\documentclass[aps,prf,reprint,notitlepage,onecolumn,superscriptaddress]{revtex4-1}
\usepackage{amsmath,amssymb}
\usepackage{graphicx}
\usepackage{subcaption}
\usepackage{bm}
\usepackage{epstopdf}
\usepackage{verbatim} % comment large blocks of text
\usepackage[normalem]{ulem}

\usepackage{xcolor}
\definecolor{light-gray}{gray}{0.5}
\definecolor{blue}{rgb}{0.0,0.0,1.0}
\definecolor{green}{rgb}{0.0,0.5,0.0}
\definecolor{red}{rgb}{1.0,0.0,0.0}
\definecolor{cyan}{rgb}{0.0,0.75,0.75}
\definecolor{magenta}{rgb}{0.75,0.0,0.75}
\definecolor{yellow}{rgb}{0.75,0.75,0.0}
\definecolor{orange}{rgb}{0.9,0.3,0.0}

\newcommand{\avg}[1]{\langle{#1}\rangle}

\newcommand{\grad}{\bm \nabla}
\newcommand{\pd}{\partial}

\newcommand{\vd}[1]{\textcolor{black}{#1}} % vassili
\newcommand{\cor}[1]{\textcolor{black}{#1}}  % kanna

\newcommand{\lhat}{\widehat}
\begin{document}

%\title{Abrupt transitions of zonal mean flow in confined two-dimensional turbulence} %a bit more general title
% \title{Abrupt reversals of zonal jets in a two-dimensional turbulent shear flow}
% \title{Bifurcations of large scale zonal jets in a two-dimensional turbulent shear flow}
\title{Transitions between turbulent states in a two-dimensional shear flow}

\author{Vassilios Dallas}
\email[]{vassilios.dallas@gmail.com}
\affiliation{Mathematical Institute, University of Oxford, Woodstock Road, Oxford OX2 6GG, UK}
\author{Kannabiran Seshasayanan}
\email[]{kannabiran.seshasayanan@gmail.com}
\affiliation{Service de Physique de l'Etat Condens\'e (SPEC), CEA Saclay B\^atiment 772, Orme des Merisiers, F-91191 Gif sur Yvette Cedex, France}
\author{Stephan Fauve}
\email[]{fauve@lps.ens.fr}
\affiliation{Laboratoire de Physique de l'\'Ecole normale sup\'erieure, ENS, Universit\'e PSL, CNRS, Sorbonne Universit\'e, Universit\'e Paris-Diderot, Sorbonne Paris Cit\'e, Paris}

\begin{abstract}
We study the bifurcations of the large scale jets in the turbulent regime of a forced shear flow using direct numerical simulations of the Navier-Stokes equations. The bifurcations are seen in the probability density function (PDF) of the largest scale mode with the control parameter being the Reynolds number based on the friction coefficient denoted as $Rh$. As one increases $Rh$ in the turbulent regime, the PDF of the large scale mode first bifurcates from a Gaussian to a bimodal behaviour, signifying the emergence of reversals of the large scale flow where the flow fluctuates between two distinct turbulent states. Further increase in $Rh$ leads to a bifurcation from bimodal to unimodal PDF which denotes the disappearance of the reversals of the largest scale mode. We attribute the latter transition to the long-time memory that the large scale flow exhibits related to low frequency $1/f^\alpha$ type of noise with $0 < \alpha < 2$. We also demonstrate that a minimal model with 15 modes, obtained from the truncated Euler equation, is able to capture the bifurcations of the large scale jets exhibited by the Navier-Stokes equations.
\end{abstract}

\maketitle


\section{Introduction} 
\label{sec:intro}

Low-frequency variability in climate exhibits recurrent patterns that are directly linked to dynamical processes of the governing dissipative system \cite{dijkstra13,ghil00}. The study of the persistence of these large scale patterns and the transitions between them plays a crucial role in understanding climate change \cite{stockeretal13}. Climate variability is often associated with dynamic transitions between different regimes, each represented by local attractors. Examples of climate phenomena where such transitions have been investigated experimentally as well as numerically are the transitions between different mean flow patterns of the Kuroshio Current in the North Pacific \cite{itohkimoto96,schmeitsdijkstra01}, the transitions between blocked and zonal flows in the midlatitude atmosphere \cite{weeksetal97}, and the transition to oscillatory behaviour in \cor{models of Quasi-biennial oscillations \cite{seminetal18,renaud2019periodicity}.}

\cor{Understanding naturally occurring bifurcations in climate systems through \vd{observational} studies is challenging due to the lack of control on the different physical processes involved. Numerical and experimental studies that model such systems can overcome such difficulties to study these bifurcations, where the parameters can be systematically controlled. They can provide important insights for our understanding of the behaviour observed in the natural counterparts. Models varying from idealised systems to the global climate models succeed in reproducing some of the observed phenomena. Such complex transition and bifurcations have also been observed in many idealised fluid mechanical systems. 
%When studying climate related phenomena one cannot control the conditions and study their effect on the behaviour of the system. On the other hand, numerical and experimental studies of such types of bifurcations, in which the parameters can be systematically controlled, can provide important insights for our understanding of the behaviour observed in the natural counterparts.
Bifurcations on turbulent flows} have been observed in a range of experimental and numerical studies for 
flow past bluff bodies \cite{wygnanskietal86,cadotetal15}, 
Rayleigh-B\'enard convection with and without rotation \cite{breuerhansen09,stevensetal09,sugiyamaetal10,chandramahendra13},
von Karman flow \cite{labbeetal96,raveletetal04,torreburguete07}, 
reversals in a dynamo experiment \cite{berhanuetal07} and 
experiments of two-dimensional turbulence \cite{micheletal16}. 
The two-dimensional turbulence experiments showed the formation of a large scale condensate which displays a series of bifurcations. These were also seen in numerical simulations that mimic the experimental set-up \cite{mishraetal15}. 
%and were captured by a statistical theory based on the truncated Euler equations \cite{shuklaetal16}.

Large scale zonal flows are found in planetary atmospheres and the ocean. Their existence motivates the study of jet formation in a turbulent flow driven at smaller scales. Anisotropy helps in the formation of these jet structures, which are usually introduced either by a $\beta$-plane that models the variation of the rotation with latitude \cite{zeitlin18}, or by a domain of non-unity aspect ratio \cite{bouchetsimonnet09}, or due to confinement and boundary conditions, see \cite{galperinread19} and references therein. \cor{These large scale jets} are known to be quite stable, evolving over longer time scales than the underlying turbulence. Such seemingly stable states undergo transitions at times \cite{bouchetetal19}, and modelling such phenomena remains an open question.

While much of the progress in fluid dynamics has been on hydrodynamic instabilities, only %\vd{a limited number of} studies 
limited progress exists on transitions/instabilities that occur when a control parameter is varied within the turbulent regime, which is the focus of this paper. 
The difficulty in studying such transitions arises from the underlying turbulent fluctuations which make any analytical progress cumbersome.
These types of transitions resemble more closely phase transitions in statistical mechanics because the instability occurs on a fluctuating background \cite{kadanoff00}. 
%So, can these turbulent fluctuations affect the scaling law of the bifurcation parameter as thermal fluctuations do for the order parameter in most phase transitions? 
% Such transitions therefore resemble closer the extreme weather transitions described above which happen between different turbulent states.
% \chk{The attempts at modelling such transitions have involved equilibrium statistical mechanics, large deviation theory and rare events algorithms \cite{bouchet2012statistical,kraichnan,shuklaetal16,bouchetetal19}. There are two different approaches using equilibrium statistical mechanics. The formalism which started with the point vortex model by Onsager has been widely studied, see \cite{bouchet2012statistical} and references therein. While the study by Kraichnan used a formalism using the truncated Euler equations (TEE). 
A recent successful attempt \cite{shuklaetal16} to model such transitions of the large scale circulation observed in experiments \cite{micheletal16} involved a statistical mechanical theory at thermal equilibrium based on the truncated Euler equation (TEE) \cite{kraichnan67,kraichnan75}.

%On that respect, the focus of this work is to study the transitions of the zonal mean flow which correspond to bifurcations on a strongly turbulent background using numerical simulations. 
%
In this paper we study the bifurcations of \cor{large scale} jets in a two-dimensional shear flow confined in the latitudinal direction. Using direct numerical simulations of the Navier-Stokes equations we quantify the behaviour of the large scale mode and the bifurcations that occur in the turbulent regime. Finally, we systematically develop a minimal model using the truncated Euler equation to capture the large scale behaviour of the Navier-Stokes equations.

\section{Problem set-up\label{sec:setup}}
\begin{figure}[!ht]
  \vspace{0.3cm}
   \includegraphics[width=0.6\textwidth]{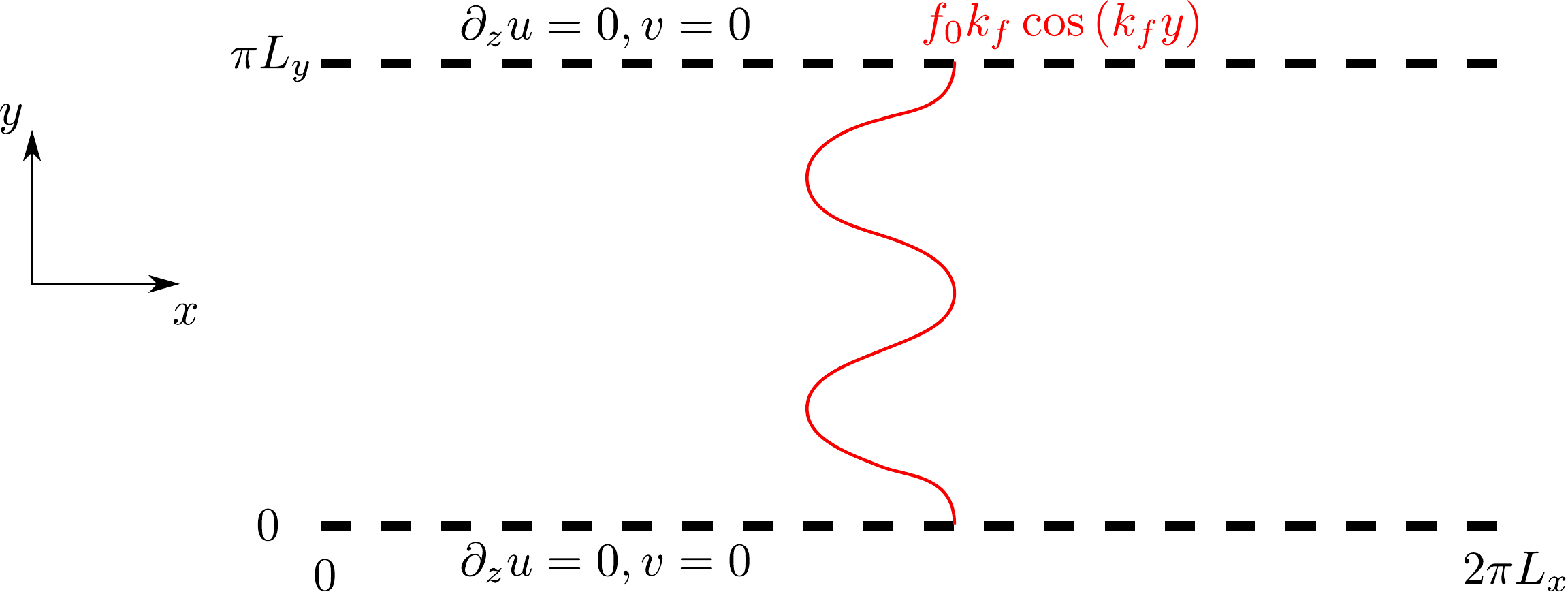}
 \caption{(Color online) Sketch of the domain under study. The red line represents the spatial form of the Kolmogorov forcing. 
 %\vd{[Kanna: Can you please change the boundary conditions in the figure to their streamfunction form and replace $\cos(k_fy) \to \cos(4y/L_y)$?]}}
 \cor{Note that $f_0 \, k_f \, \cos(k_f y)$ profile corresponds to the force that acts on \vd{$u$, the $x$ component of the velocity field,} obtained from taking the $y$-derivative of equation \eqref{eq:NS}.} }
 \label{fig:domain}
 \end{figure}

We consider the two-dimensional Navier-Stokes equations for an incompressible velocity field ${\bf u} =  \grad \times \psi{\bf \hat z}$ forced by a Kolmogorov type forcing in an anisotropic domain 
%of dimensions 
$(x,y) \in \left[0, 2 \pi L_x \right] \times \left[0, \pi L_y \right]$ as illustrated in Fig. \ref{fig:domain}. The governing equation written in terms of the streamfunction $\psi (x,y,t)$ is given by,
\begin{equation}
 \pd_t \psi + \grad^{-2}\{\grad^2\psi,\psi\} = \nu\grad^2\psi - \mu\psi + f_0\sin(k_f y)
 \label{eq:NS}
\end{equation}
where $\{f,g\} = f_x g_y - g_x f_y$ is the standard Poisson bracket (subscripts here denote differentation), $\nu$ is the kinematic viscosity, $\mu$ is the friction coefficient, $f_0$ is the amplitude of the Kolmogorov forcing and \cor{$k_f$ is the number of half wavelengths in the forcing. Note that the profile shown in Fig. \ref{fig:domain} denotes the form of the forcing as it appears in the $x$-component of the momentum equation. Since $u = \partial_y \psi$, it implies that the forcing written in Eq. \eqref{eq:NS} becomes $f_0 k_f \cos \left( k_f \, y \right)$ in the governing equation for $u \left( x, y, t \right)$. The forcing shown in Fig. \ref{fig:domain} corresponds to $k_f L_y = 4$. }
%
%Our choice of the external driving force can be physically justified. In an atmospheric model the Kolmogorov forcing $f_0\sin(k_f y)$ can represent the transfer of angular momentum into midlatitudes due to the tropical Hadley cell \cite{lorenz67}. In the case of the ocean, the forcing term is the curl of the wind stress $\grad \times \tau = f_0\sin(k_f y)$. A wind stress of the form $\tau = f_0/k_f (\cos(k_f y),0)$ mimics the annually averaged zonal wind distribution over the North Atlantic and North Pacific with westward winds over the mid-latitudes and eastward winds in the tropics and polar latitudes \cite{dijkstraetal15}.
The boundary conditions are taken to be periodic in the $x$
%(or zonal) 
direction and free-slip in the $y$
%(or latitudinal) 
direction, i.e. $\psi_{yy} = \psi_x = 0$ at $y = 0, \pi L_y$. 
We are interested on the effect of the friction coefficient $\mu$ on the dynamics of large scale jets. Thus, we vary $\mu$ by keeping fixed the ratio $\nu/(f_0^{1/2}L_x) = 10^{-3}$, \cor{the number of half-wavelengths with respect to the height which gives} $k_f L_y = 4$ and the aspect ratio of the domain $2 \pi L_x/(\pi L_y) = 2$.
For the rest of the article, all quantities are non-dimensionalised with the rms velocity $u_{rms} = \avg{|{\bf u}|^2}^{1/2}$, the length scale $L_x$ and the time scale $L_x/u_{rms}$. Here, the angle brackets $\avg{.}$ denote integration over the domain and time. Then, the Reynolds number can be defined as $Re = u_{rms} L_x /\nu$ and the friction Reynolds number as 
\begin{equation}
Rh = u_{rms}/(\mu L_x),
\end{equation}
which is the ratio of the inertial term to the friction term in Eq. \eqref{eq:NS}. %\ks{In the rest of the article, all quantities are non-dimensionalised.}

We perform direct numerical simulations (DNS) by integrating Eq. \eqref{eq:NS} using the pseudospectral method  \cite{gottlieborszag77}. We decompose the streamfunction into basis functions with Fourier modes in the $x$ direction and Sine modes in the $y$ direction that satisfy the boundary conditions,
\begin{equation}
 \psi(x,y,t) = \sum_{k_x=-\frac{N_x}{2}}^{\frac{N_x}{2}-1} \sum_{k_y=1}^{N_y} \widehat{\psi}_{_{k_x,k_y}}(t) e^{i k_x x} \sin(k_y y),
 \label{eq:decomp_FS}
\end{equation}
with $\widehat{\psi}_{_{k_x,k_y}}$ being the amplitude of the mode $\left(k_x, k_y \right)$ and $(N_x, N_y)$ denote the number of grid points in the $x, y$ coordinates respectively.
A third-order Runge-Kutta scheme is used for time advancement and the aliasing errors are removed with the two-thirds dealiasing rule which implies that the maximum wavenumbers are $k_x^{max} = N_x/3$ and $k_y^{max} = 2N_y/3$. 
The resolution was fixed to $(N_x,N_y) = (512,128)$ for all the simulations done in this study. Note that our resolution is limited in this study to be able to integrate for very long times that are required to accumulate reliable statistics for the large scale flow transitions. We also verified the statistics at higher resolutions for certain parameter values of $Rh$.

%
%\section{Large scale mean flow transitions}
\vd{\section{Large scale flow transitions}}

We are interested in quantifying the transitions of the large scale flow as a function of the control parameter $Rh$. 
\cor{The large scale mean flow is defined using the amplitude of the largest mode $\widehat{\psi}_{0,1} \left( t \right)$ as} 
\vd{
\begin{equation}
 U(y,t) = \widehat{\psi}_{0,1}(t) \cos \left( y \right) \hat{\bf e}_x,
\end{equation}
where
\begin{equation}
 \widehat{\psi}_{0,1}(t) = \frac{1}{2\pi^2}\int_0^{2\pi }\int_0^{\pi } \psi(x,y,t) \, \sin(y) \, dy dx.
\end{equation}
This large scale flow is a shear flow along the $x$ direction with zero mean value. Fig. \ref{fig:domain} shows} \cor{that the forcing and the boundary conditions are symmetric with respect to the centreline $y = \pi/2$. The mean flow on the other hand breaks the centreline symmetry whenever $\widehat{\psi}_{0,1}$ is non-zero.} %Due to the symmetry of the forcing and the domain with respect to the centreline $y = \pi L_y/2$, a non-zero value of $\widehat{\psi}_{0,1}(t)$ does not respect the symmetry with respect to the centreline. 
For $Rh \ll 1$, the forcing drives a laminar flow (see the red profile shown in Fig. \ref{fig:domain}) which essentially consists of two westward and two eastward jets, with a zero projection onto the largest mode in the system ($\widehat{\psi}_{0,1} = 0$). 
This laminar flow becomes unstable \cite{thess92} above a critical value of \cor{$Rh = Rh_1^{c} \approx 0.697$ leading to a 
degenerate Hopf bifurcation which we report in detail \vd{in} \cite{sdf19b}}. This saturated state above \cor{$Rh_1^{c}$} still has zero projection onto the largest mode. As $Rh$ increases even further the system undergoes another Hopf bifurcation at \cor{$Rh = Rh_2^{c} \approx 1.165$} with the largest mode directly forced due to the nonlinearity \vd{\cite{sdf19b}}. Then, for \cor{$Rh > Rh_2^{c}$} the system undergoes a sequence of bifurcations, similar to \cite{fauveetal17}, until it reaches a 2D turbulent state. In this study we focus on the turbulent regime where we show that there are bifurcations in the behaviour of the large scale flow on top of a turbulent background. 

\begin{figure}[!ht]
 \begin{subfigure}{0.49\textwidth}
   \includegraphics[width=\textwidth]{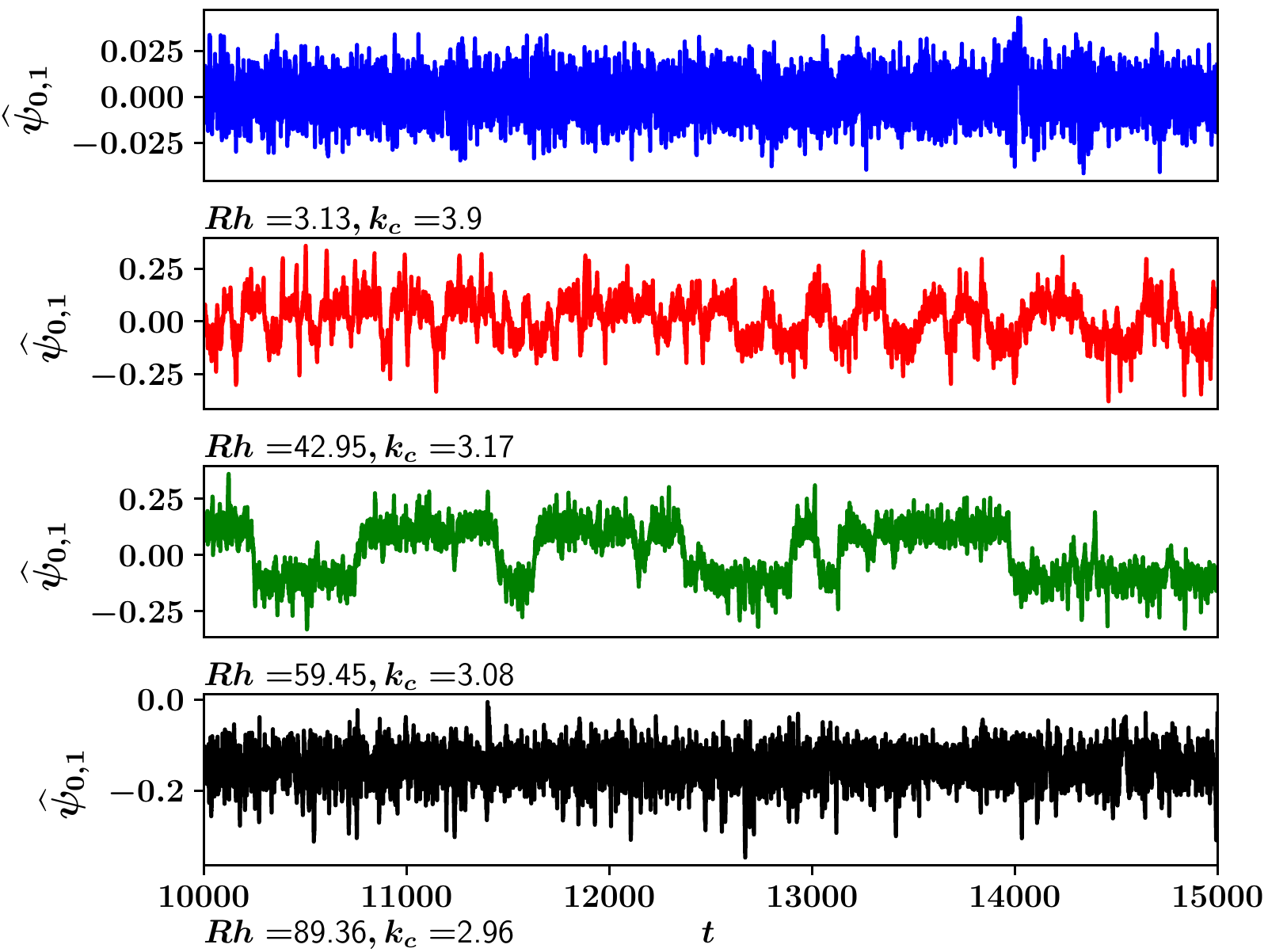}
   \caption{}
\label{fig:dns_time_series}
 \end{subfigure} 
 \begin{subfigure}{0.49\textwidth}
   \includegraphics[width=\textwidth]{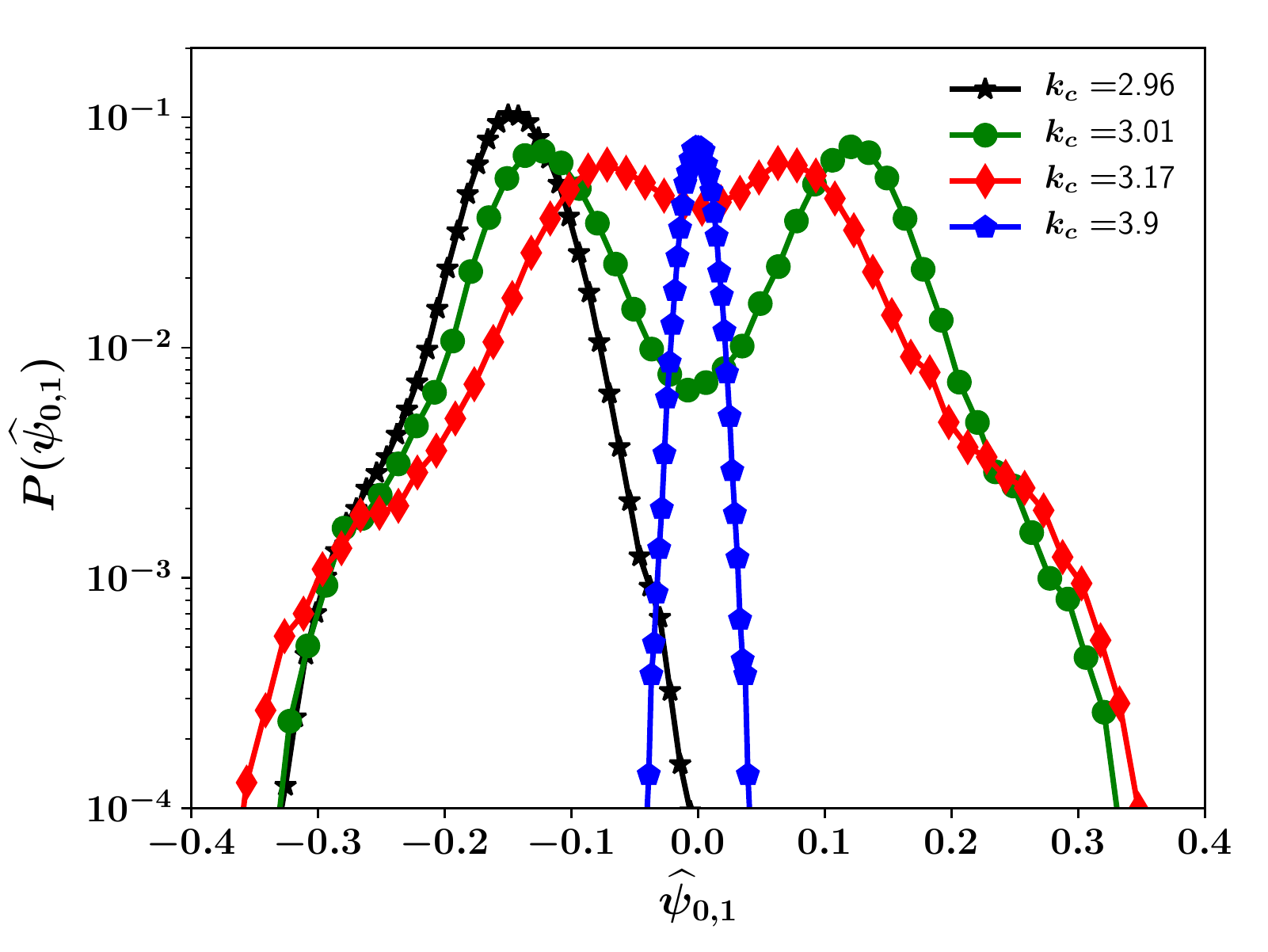}
   \caption{}
\label{fig:pdf_NS}
 \end{subfigure}
  \caption{(Color online) (a) Time series of the large scale mode $\lhat\psi_{0,1}$ from the DNS and their corresponding (b) PDFs for different values of $Rh$}
 \label{fig:dns} 
 \end{figure}

In the turbulent regime $Rh$ still plays the role of the bifurcation parameter but in this case the relevant observables are statistical quantities. In Fig. \ref{fig:dns_time_series} we show the time series of $\widehat{\psi}_{0,1}$ and in Fig. \ref{fig:pdf_NS} their corresponding PDFs for different values of $Rh$.
At $Rh = 3.13$, the time series is turbulent with the amplitude of $\widehat{\psi}_{0,1}$ fluctuating randomly around the zero mean implying that the symmetry with respect to the centreline is restored for the mode in a statistical sense. The PDF of this time series is close to Gaussian. 
For $Rh = 42.95$ we start to see longer duration of time where $\lhat\psi_{0,1}$ has a non-zero amplitude, the corresponding PDF is bimodal with two distinct peaks. This represents the first bifurcation where the system transitions from a Gaussian distribution to a bimodal distribution with two symmetric maxima. This behaviour is related to the emergence of two symmetric states where the time series is characterised by abrupt and random reversals between these two states. One state is when %the zonal mean flow points westward 
$\lhat\psi_{0,1} > 0$ and the other symmetric state is when %it points eastward 
$\lhat\psi_{0,1} < 0$ for long duration.
For $Rh = 59.45$ we get random reversals of the large scale flow with a PDF which is bimodal with more distinct peaks. As we keep increasing $Rh$ the reversals become rarer until we get to a state where there are no more large scale flow reversals observed indicating another bifurcation. This is seen in the case with $Rh = 89.36$ where the system was never observed to reverse and the corresponding PDF is unimodal with a non-zero mean. \cor{The PDF for the largest $Rh = 89.36$ chooses either \vd{a} positive or negative value of $\widehat{\psi}_{0,1}$ depending on the initial condition. \vd{The unimodal distribution indicates that ensemble averaging is not equivalent to the time averaged system.}} 
\cor{For each value of $Rh$ examined we have also calculated the corresponding value of $k_c$ which is defined by,
\begin{equation}
\cor{k_c = (\Omega/E)^{1/2},}
\label{eq:kc_eq}
\end{equation}
\vd{where the kinetic energy and the enstrophy are defined respectively, 
\begin{equation}
 E = \avg{|\grad \psi|^2}_{_A}, \quad  \Omega = \avg{|\grad^2 \psi|^2}_{_A},
 \label{eq:invariants}
\end{equation}
with the angular brackets $\avg{\cdot}_{_A}$ denoting integration over $x$ and $y$.}
\cor{The value of $k_c$ is later used to compare with the truncated Euler system, where $k_c$ acts as the control parameter \vd{of the system} instead of $Rh$.}}

\cor{A similar sequence of bifurcations of the large scale circulation has been observed in laboratory experiments of quasi-2D flow \cite{micheletal16} and also in numerical simulations of 2D turbulence \cite{shuklaetal16}. In those studies the boundary conditions were no-slip or free-slip in both $x, y$ boundaries. The energy spectra from our DNS (not shown here) are found to have a peak at an intermediate wavenumber $k = 2$ when the system displays bimodal behaviour of $\widehat{\psi}_{0,1}$. Here $k = \sqrt{k_x^2 + k_y^2}$ is the \vd{amplitude} of the wavevector. The $2D$ turbulent flow is not in the condensate regime where the large scale mode is much larger in amplitude when compared to the rest of the modes. This is in contrast to the study \cite{shuklaetal16} where those DNS spectra peaked at the lowest wavenumber $k_{min} = 1$ in the bimodal regime.}

\cor{To understand the flow structure in the different regimes we \vd{analyse the} mean flow profile $U(y, t)$ defined as
\begin{align}
U \left( y, t \right) = \frac{1}{2 \pi} \int_{0}^{2 \pi} u \left( x, y, t \right) dx. 
\end{align}
Figure \ref{fig:mean_flow} illustrates the mean velocity profile for the flow with $Rh = 59.45$. 
%(see also the times series of $\lhat\psi_{0,1}$ for this flow in Fig. \ref{fig:dns_time_series}).
 \begin{figure}[!ht]
 \begin{subfigure}{0.49\textwidth}
   \includegraphics[width=\textwidth]{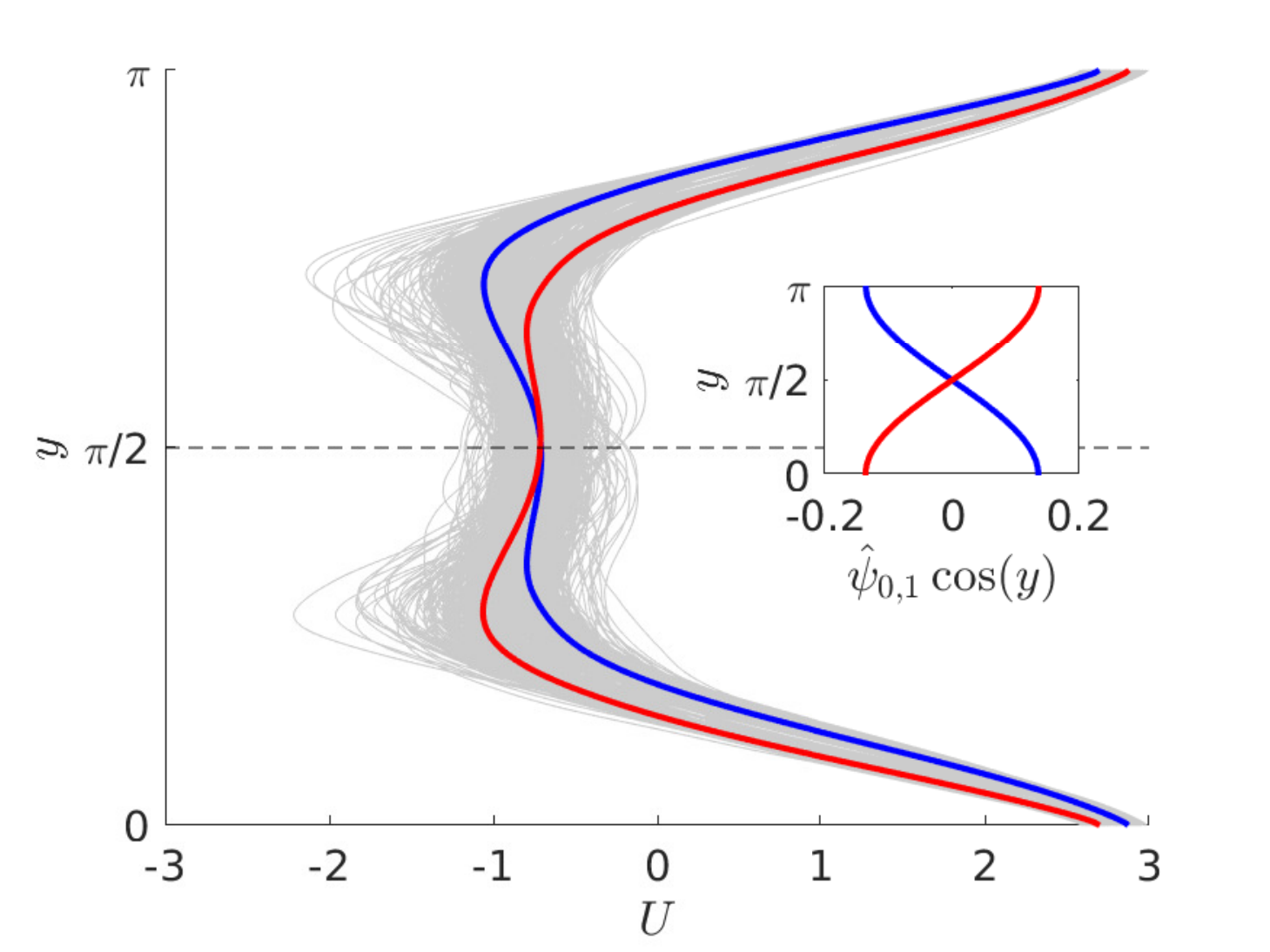}
   \caption{}
 \label{fig:mean_flow} 
 \end{subfigure} 
  \begin{subfigure}{0.49\textwidth}
   \includegraphics[width=\textwidth]{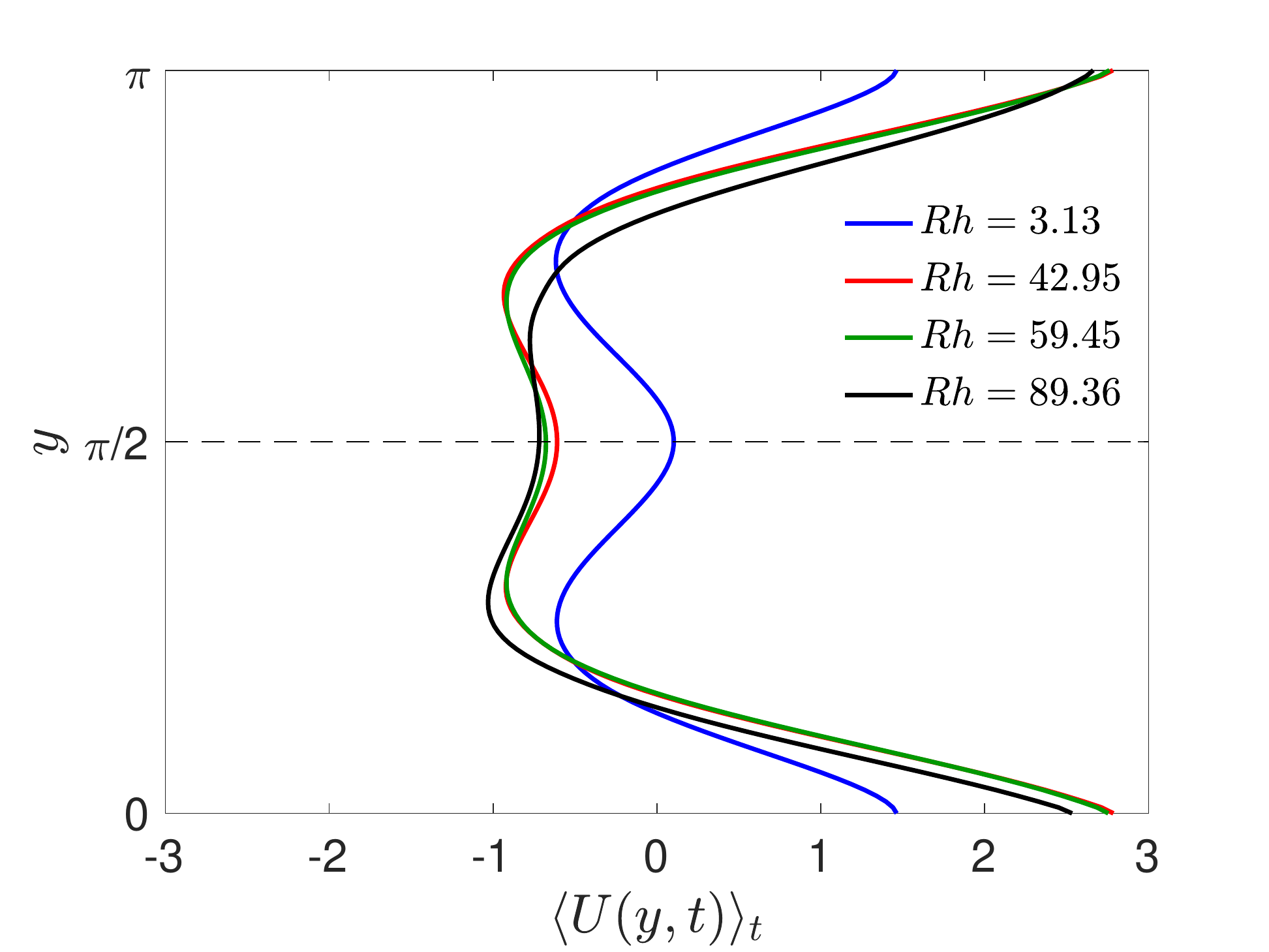}
   \caption{}
\label{fig:diff_mean}
 \end{subfigure}
 \caption{(Color online) \cor{ (a) Conditionally time averaged mean flow profiles when $\lhat\psi_{0,1} > 0$ (blue) and $\lhat\psi_{0,1} < 0$ (red) for $Rh = 59.45$. Grey curves represent the instantaneous mean flow profiles. The inset shows the projection of the conditionally time averaged mean flow profiles onto the large scale mode. (b) The time averaged mean flow profiles $\left\langle U ( y , t ) \right\rangle_t$ for four different values of $Rh$ as mentioned in the legend. } }
 \label{fig:means} 
 \end{figure} 
The blue and red curves represent the mean flow profiles which have been time averaged over the time intervals conditional on $\lhat\psi_{0,1} > 0$ and $\lhat\psi_{0,1} < 0$, 
respectively. A part of these intervals is shown in the time series for $Rh = 59.45$ in Fig. \ref{fig:dns_time_series}. The grey curves indicate the instantaneous realisations of the mean flow profile at different times. The conditionally time-averaged mean flow profiles (blue and red curves in Fig. \ref{fig:mean_flow}) are then projected onto the largest mode, 
which are shown in the inset of Fig. \ref{fig:mean_flow}. When $\lhat\psi_{0,1} > 0$ the conditionally time averaged profile, blue curve in Fig. \ref{fig:mean_flow}, shows a stronger westward jet in the upper half of the domain ($\pi/2 < y < \pi$) than in the lower half ($0 < y < \pi/2$). The reverse happens for $\lhat\psi_{0,1} < 0$ as seen from the red curve in Fig. \ref{fig:mean_flow}. The transition between $\lhat\psi_{0,1} > 0$ to $\lhat\psi_{0,1} < 0$ state captures the meandering of the jets from the upper to the bottom half of the domain. Thus, reversals of the large scale mode leads to modulations or meandering of the jets which occurs on a time scale much longer than the eddy turn over time.}

\cor{Fig. \ref{fig:diff_mean} shows the time averaged mean flow profile $\left\langle U (y, t ) \right\rangle_t$ for four different values of $Rh$. The lowest value of $Rh = 3.13$ where the PDF of the large scale mode $\widehat{\psi}_{0,1}$ is Gaussian shows a mean flow profile which \vd{resembles} the forcing. %The time averaged flow for the lowest $Rh$ 
\vd{This time averaged profile} is symmetric about the centreline $y = \pi/2$ while instantaneous profiles break the centreline symmetry. As we increase $Rh$, in the regime where the PDF of $\widehat{\psi}_{0,1}$ is bimodal, we see that the mean flow profile deviates from the form of the forcing and \vd{two westward jets} develops at the centre. The centreline symmetry is still respected by the time averaged mean flow profile. For the largest value of $Rh$ shown in Fig. \ref{fig:diff_mean} which corresponds to a \vd{unimodal} PDF of $\widehat{\psi}_{0,1}$, the time averaged mean profile breaks the centreline symmetry. Note that the bifurcations of the large scale mode $\widehat{\psi}_{0,1}$ are related to changes in the form of the time averaged mean flow $\left\langle U(y, t) \right\rangle_t$.} \\

\section{Minimal model: truncated Euler equation}
Now we seek for a minimal model to capture the dynamics of the random transitions of the large scale flow. To do this we consider the incompressible Euler equation truncated at a maximum wavenumber $k_{max}$ using a %circular 
Galerkin truncation, which in 
Fourier-Sine basis take the form
\begin{equation}
 \pd_t \lhat\psi_{\bm k} = \sum_{\bm p, \bm q} A_{\bm k,\bm p,\bm q} \lhat\psi_{\bm p} \lhat\psi_{\bm q}
 \label{eq:tee}
\end{equation}
with the interaction coefficients to be given by 
\begin{align}
A_{\bm k,\bm p,\bm q} = \frac{i}{2}(q^2 - p^2)k^{-2}\delta_{k_x,p_x + q_x}[ (p_xq_y - p_yq_x) \delta_{k_y,p_y + q_y} 
                                                       +(p_x q_y + p_y q_x)                                                     (\delta_{k_y,p_y - q_y} - \delta_{k_y,q_y - p_y})],
 \label{eq:tee2}                                                       
\end{align}
where %$\bm p \times \bm q = p_xq_y - p_yq_x$ and
$\delta_{i,j}$ stands for the Kronecker delta. 
\cor{To derive Eqs. \eqref{eq:tee}, \eqref{eq:tee2} \vd{one has to replace Eq. \eqref{eq:decomp_FS}, the Fourier-Sine decomposition of $\psi (x, y, t)$, into the governing equation Eq. \eqref{eq:NS}.} Taking the inverse Fourier transform of the resulting equation we end up with Eq. \eqref{eq:tee} and the expression for the interaction coefficients in Eq. \eqref{eq:tee2}. 
\vd{This} truncated system of ODEs} conserves exactly the two quadratic invariants of the Euler Equation (i.e. Eq. \eqref{eq:NS} with the RHS equal to zero), \vd{the kinetic energy $E$ and the enstrophy $\Omega$ (see definitions in Eq. \eqref{eq:invariants}).}

The fundamental idea of equilibrium statistical mechanics is to construct a probability density on the phase space of a dynamical system based only on the invariants of the dynamics. It is well known \cite{lee52,kraichnan67,kraichnan75} that the probability density $\mathcal P(\lhat\psi_{\bm k},t)$ of all the amplitudes of the modes $\lhat\psi_{\bm k}$ in the $N$-dimensional phase space of the Euler equation obeys Liouville's equation 
\cor{
\begin{equation}
\pd_t \mathcal P + \sum_k \dot{\lhat{\psi_{\bm k}}}\pd_{\lhat\psi_{\bm k}}\mathcal P = 0, 
\end{equation}
}
where the sum here is over all $N$ modes $\lhat\psi_{\bm k}$. 
\cor{Here the volume in phase space is conserved by the dynamics according to Eq. \eqref{eq:tee} and therefore the microcanonical formalism can be used.} 
In the microcanonical formalism \cite{khinchin60}, $\mathcal P(\lhat\psi_{\bm k},t)$ is uniform on all the states with a given set of values for the invariants and vanishes for any other set of values. In our case, if one takes into account only the quadratic invariants of the Euler equation, the microcanonical density is defined as
\begin{equation}
% \mathcal P = Z^{-1}\exp(-\alpha E - \beta \Omega),
\mathcal P = Z(E,\Omega)^{-1}\delta(\lhat E - E)\delta(\lhat \Omega - \Omega).
\end{equation}
where $\lhat E$ and $\lhat \Omega$ are the energy and enstrophy values at absolute equilibrium \cite{kraichnan75} and the normalisation factor $Z(E,\Omega)$, known as the structure function of the system \cite{khinchin60}, measures the surface of constant energy and enstrophy over the phase space volume elements $d\Gamma = \prod_{\bm k} d\lhat\psi_{\bm k}$. This distribution is appropriate for describing an isolated system of initial energy $E$ and initial enstrophy $\Omega$, with no exchange with its surroundings, under the assumption that the system exhibits suitable ergodic properties \cite{khinchin60}.

\begin{figure}[!ht]
 \begin{subfigure}{0.49\textwidth}
   \includegraphics[width=\textwidth]{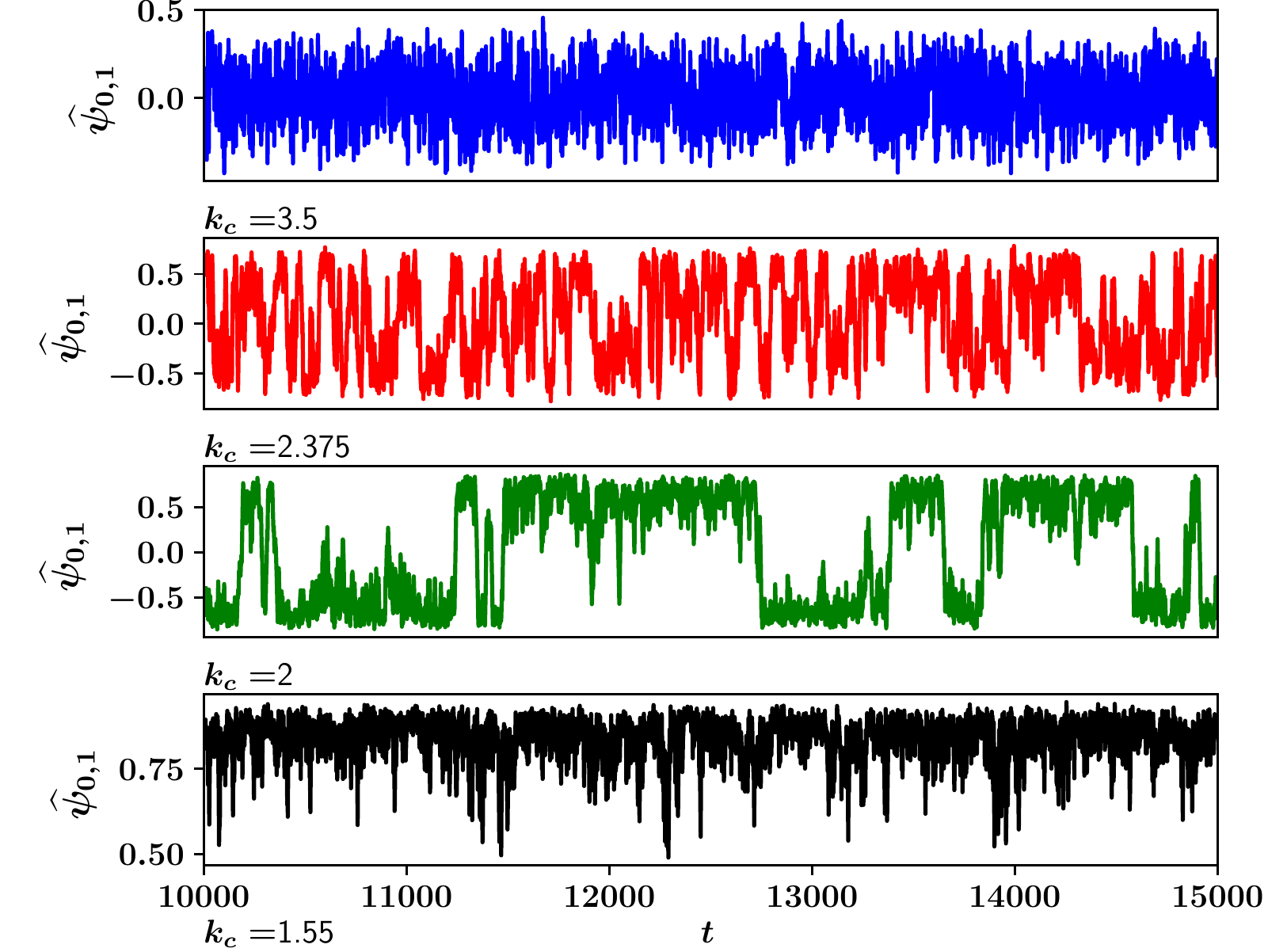}
   \caption{}
\label{fig:euler_time_series}
 \end{subfigure} 
 \begin{subfigure}{0.49\textwidth}
   \includegraphics[width=\textwidth]{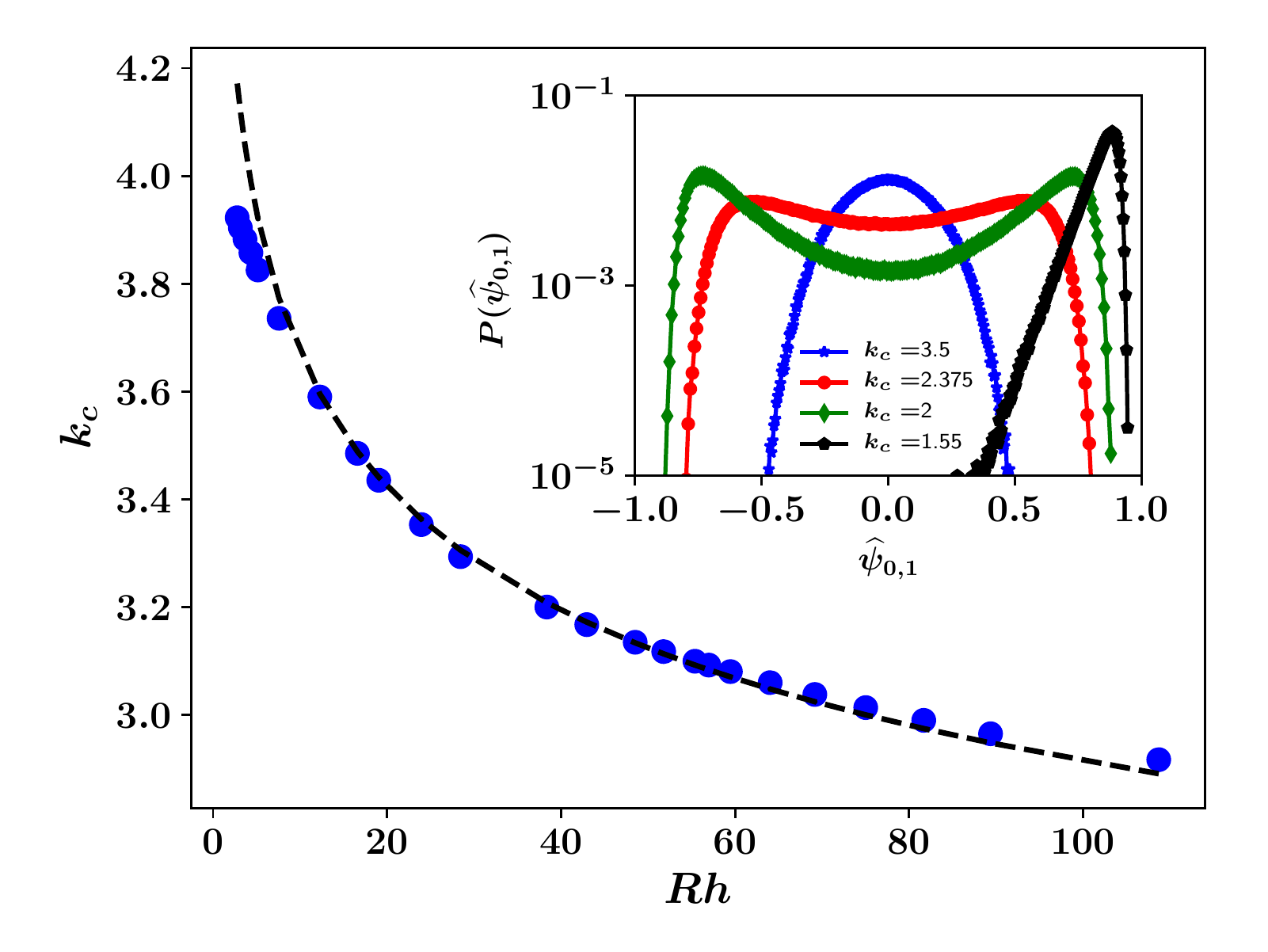}
   \caption{}
\label{fig:kc_Rh}
 \end{subfigure}
  \caption{(Color online) (a) Time series of the large scale mode $\lhat\psi_{0,1}$ for the TEE system. (b) $k_c = (\Omega/E)^{1/2}$ as a function of $Rh$ from the DNS. The dashed line is the fit from Eq. \eqref{eq:kcRh}. Inset: PDFs of the time series of $\lhat\psi_{0,1}$ for the TEE system. }
  \label{fig:tee} 
 \end{figure}

\cor{For the truncated Euler equation (TEE) the control parameter depends only on the initial conditions. The wavenumber $k_c$, which is the square root of the ratio of initial enstrophy to initial energy \vd{(see Eq. \eqref{eq:kc_eq})}, is taken as the control parameter for the TEE system.} %, where $E$ and $\Omega$ in this case take the values at the initial conditions. 
In the full Navier-Stokes equations the equivalent wavenumber is controlled by the friction coefficient (i.e. the value of $Rh$) when the viscosity is fixed. 
In Fig. \ref{fig:kc_Rh} we show the dependence of $k_c$ on $Rh$ for the Navier-Stokes equations and we see that as $Rh$ increases the wavenumber $k_c$ decreases like 
\begin{equation}
 k_c \propto Rh^{-1/10}. 
 \label{eq:kcRh} 
\end{equation}
This fit comes from the observation that the energy scales like $E \propto Rh^{4/5}$ and the enstrophy scales like $\Omega \propto Rh^{3/5}$ for the values of $Rh$ in a bigger range than the bimodal regime. So far, we do not have a theoretical explanation for these scalings. %and for this reason we decided not to present them here.

The initial conditions for the TEE are chosen such that $k_c$ is as close as possible to the DNS results. %and within the range $k_\alpha^2 < k_c^2 < k_\beta^2$ to make sure we fall in regime II. 
We then integrate Eqs. \eqref{eq:tee}-\eqref{eq:tee2} using a numerical scheme similar to the one used for the Navier-Stokes equations (see section \ref{sec:setup}). The Fourier amplitudes for the TEE satisfy $\lhat\psi_{\bm k} = 0$ for $|k_x| > k_x^{max}$ and $|k_y| > k_y^{max}$. Note that $k_x^{max}$ and $k_y^{max}$ are our free parameters for the truncation of the Euler equation. 
The starting point to obtain a minimal model from the TEE was first to set $k_x^{max} = k_y^{max} = 4$, which corresponds to the value of the forcing wavenumber $k_f$ we chose for the Navier-Stokes equations. The basic principle behind this choice is that the dynamics of the large scales, i.e. $k < k_f$, are not affected by viscosity and hence they are governed by the dynamics of the Euler equation as it has been demonstrated in previous studies \cite{dfa15,ld16,shuklaetal16,alexakisbrachet19}. \cor{We then \vd{removed} one by one the high wavenumber modes and tried to reproduce the bifurcations as we varied the control parameter $k_c$. Using this procedure we ended up with a minimal model of $15$ complex amplitude modes that captures qualitatively the different bifurcations of $\widehat{\psi}_{0,1}$ similar to ones observed in the Navier-Stokes equations. The $15$ mode minimal model is composed of the following complex amplitude modes,
\vd{
\begin{align}
\widehat{\psi}_{0,1},\; 
\widehat{\psi}_{0,2},\; 
\widehat{\psi}_{0,3},\;
\widehat{\psi}_{0,4},\; 
\widehat{\psi}_{1,1},\; 
\widehat{\psi}_{1,2},\;
\widehat{\psi}_{1,3},\;
\widehat{\psi}_{1,4},\; 
\widehat{\psi}_{2,1},\;
\widehat{\psi}_{2,2},\; 
\widehat{\psi}_{2,3},\; 
\widehat{\psi}_{2,4},\;
\widehat{\psi}_{3,1},\; 
\widehat{\psi}_{3,2},\; 
\widehat{\psi}_{3,3}.
\end{align}
}
Only the positive $k_x$ modes are used in the counting of $15$ modes, the negative $k_x$ modes are directly related by $\widehat{\psi}_{-k_x, k_y} = \widehat{\psi}_{k_x, k_y}^{*}$. A further truncated model of $11$ modes could reproduce large scale flow reversals but the solution starts to depend on the initial conditions. So, we use the $15$ mode model to study the bifurcations of the large scale mode $\widehat{\psi}_{0,1}$. }
% After a systematic truncation of the Euler equation, we find that the minimal model which captures qualitatively the bifurcations between the different turbulent regimes observed for the Navier-Stokes equations consists of 15 ODEs for the complex large scale modes, while the minimal model that can produce large scale mean flow reversals but not the bifurcations consists of 11 ODEs.

In Fig. \ref{fig:euler_time_series} we show the time series of the amplitude of the large scale mode $\lhat\psi_{0,1}$ for different values of $k_c$ for the $15$ mode minimal model and the inset of Fig. \ref{fig:kc_Rh} shows the corresponding PDF distributions of the time series.
For $k_c = 3.5$ the TEE system exhibits a turbulent time series with a Gaussian PDF. At $k_c = 2.375$ the system shows abrupt and random reversals of the large scale flow with a bimodal PDF. Thus, when the system transitions from a Gaussian to a bimodal distribution, we observe the first bifurcation over a turbulent background between these two values of $k_c$. As $k_c$ decreases, the bimodality in the distribution becomes more pronounced and the reversals become less frequent. 
Finally, for $k_c = 1.55$ there is no longer any reversal for the duration of the simulation that lasted tens of thousands of turnover times. Here, the large scale mode bifurcates to a one-sided unimodal PDF.
Clearly this minimal set of ODEs captures the bifurcations between the different turbulent regimes observed in the full Navier-Stokes equations. 

In the simulations we observe two different scenarios of emergence and disappearance of large scale flow reversals. At high values of $k_c$ (or low $Rh$ values for Navier-Stokes), large scale flow reversals are absent because the time series are very fluctuating and one can no longer identify the two states. At low $k_c$ (or high $Rh$ for Navier-Stokes), mean flow reversals become less and less probable and eventually are no longer observed in the duration of the simulation. 
These two scenarios of emergence and disappearance of reversals have also been observed in other contexts \cite{heraultetal15b,shuklaetal16,fauveetal17,pereiraetal19} and hence we believe that they are generic.

\section{Long-time memory and $1/f^\alpha$ noise \label{sec:1/f}}

\begin{figure}[!ht]
   \includegraphics[width=0.5\textwidth]{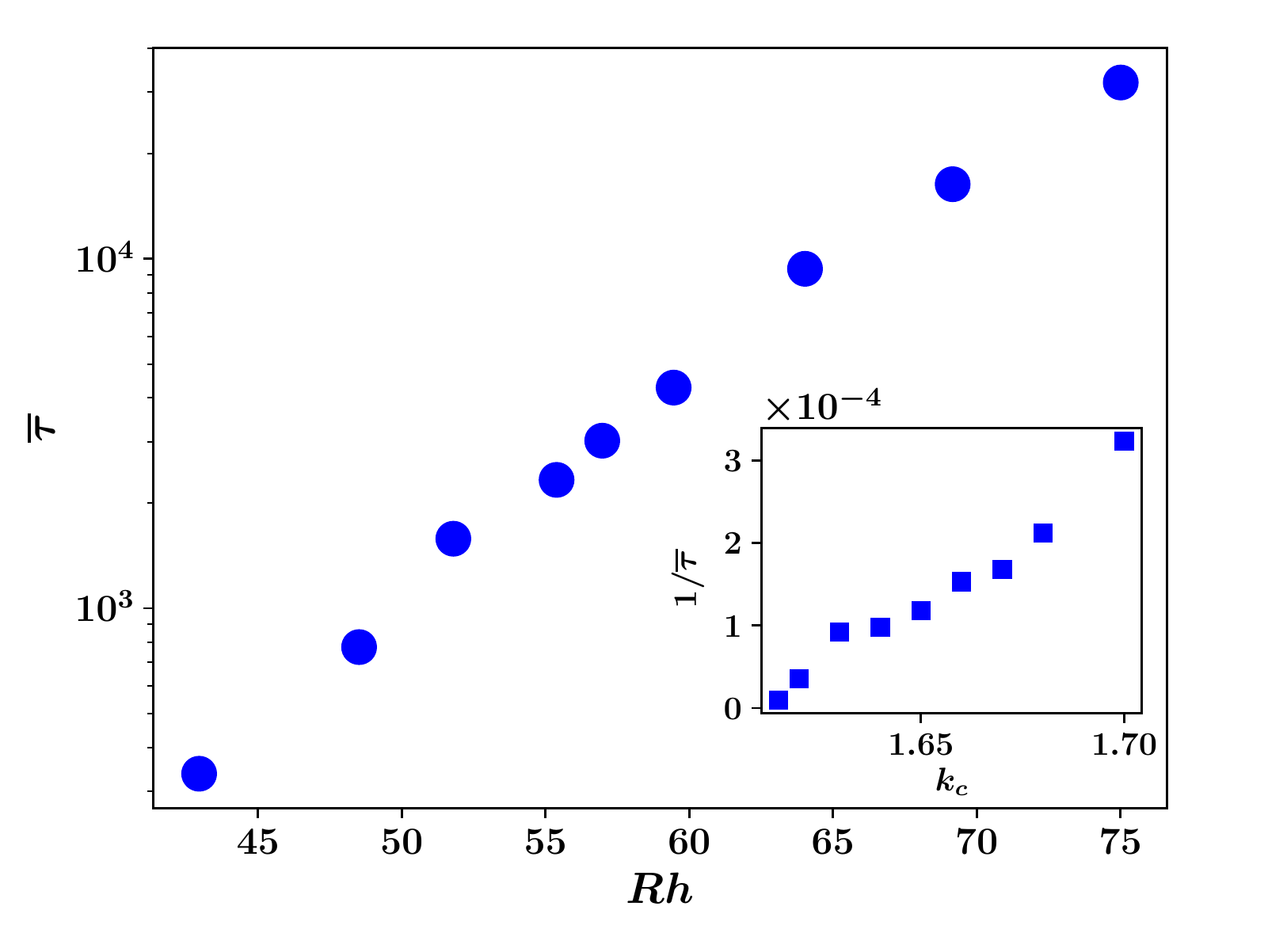}
 \caption{(Color online) Log-linear plot of the mean waiting time $\overline{\tau}$ between successive reversals as a function of $Rh$ in the bimodal regime of the Navier-Stokes equations. Inset: Plot of the mean waiting frequency $\overline{\tau}^{-1}$ between successive reversals as a function of $k_c$ in the bimodal regime of the TEE.} 
 \label{fig:tau_Rh} 
 \end{figure}

A quantity of interest is the mean waiting time $\overline{\tau}$ between successive reversals of $\lhat\psi_{0,1}$ for both the Navier-Stokes equations and the minimal model of the TEE in the bimodal regime. Here the bar denotes average over all waiting times.
Figure \ref{fig:tau_Rh} shows $\overline{\tau}$ as a function of $Rh$ for the Navier-Stokes equations and the inset shows how $1/\overline{\tau}$ varies as a function of $k_c$ for the TEE.
Figure \ref{fig:tau_Rh} shows an exponential behaviour, %of $\tau$ with $Rh$ 
i.e. 
\begin{equation}
 \log(\overline{\tau}) \propto Rh 
\end{equation}
for the Navier-Stokes equations, while a power-law behaviour 
\begin{equation}
 \overline{\tau} \propto (k_c - k_c^{crit})^{-1} 
\end{equation}
with a critical wavenumber $k_c^{crit} \simeq 1.55$ for the minimal model (see inset of Fig.  \ref{fig:tau_Rh}). For $k_c < k_c^{crit} = \sqrt{2}$ large scale flow reversals are not possible in the TEE minimal model. This can be obtained by expanding the energy and enstrophy as sums of the Fourier modes, viz.
\begin{align}
E &=  \avg{|\grad \psi|^2} = \sum_{k_x = -N_x/2}^{N_x/2} \sum_{k_y = 1}^{N_y} (k_x^2 + k_y^2) |\lhat\psi_{k_x,k_y}|^2 
\label{eq:E} \\
\Omega &=  \avg{|\grad^2 \psi|^2} = \sum_{k_x = -N_x/2}^{N_x/2} \sum_{k_y = 1}^{N_y} (k_x^2 + k_y^2)^2 |\lhat\psi_{k_x,k_y}|^2
\label{eq:Omega}
\end{align}
which in short can be written as $E = (|\lhat\psi_{0,1}|^2 + 2 |\lhat\psi_{1,1}|^2 + ...)$ and $\Omega = (|\lhat\psi_{0,1}|^2 + 4 |\lhat\psi_{1,1}|^2 + ...)$. \cor{Due to conservation of both energy and enstrophy, $k_c$ is always fixed to its initial value. Starting with $k_c^2 < 2$ with a non-zero initial value of $\widehat{\psi}_{0,1}$, if $\lhat\psi_{0,1} \left( t \right) = 0 $ at some instant $t > 0$, \vd{then Eqs. \eqref{eq:E} and \eqref{eq:Omega} imply $k_c^2 = \Omega / E \geq 2$,} which can only occur if the conservation of energy and enstrophy \vd{is} broken. This cannot happen in the TEE model, thus for $k_c^2 < 2$ the large scale mode satisfies the condition $\lhat\psi_{0,1} \left( t \right) \neq 0$ at all times implying no reversal of $\widehat{\psi}_{0,1}$.} In contrast, the Navier-Stokes equations do not involve conserved quantities that prevent reversals, even for $Rh \gg 1$. In addition, all the wavenumbers $k > k_f$ that are suppressed in the truncated Euler model can act as an additional source of noise for the Navier-Stokes equations and trigger reversals. We believe that these are the reasons why we observe a different behaviour between the Navier-Stokes equations and the TEE. Note that when we plot $\overline{\tau}$ as a function of $k_c$ for the Navier-Stokes equations we observe $\overline{\tau} \propto \exp(k_c^{-10})$ (not shown here), which is in agreement with $k_c \propto Rh^{-1/10}$ observed in Fig. \ref{fig:kc_Rh}. Both the exponential and the power-law behaviour suggest long-term dynamical memory effects of the large scale flow as the bifurcation parameters are varied within the bimodal regimes of the two systems.

\begin{figure}[!ht]
 \begin{subfigure}{0.49\textwidth}
   \includegraphics[width=\textwidth]{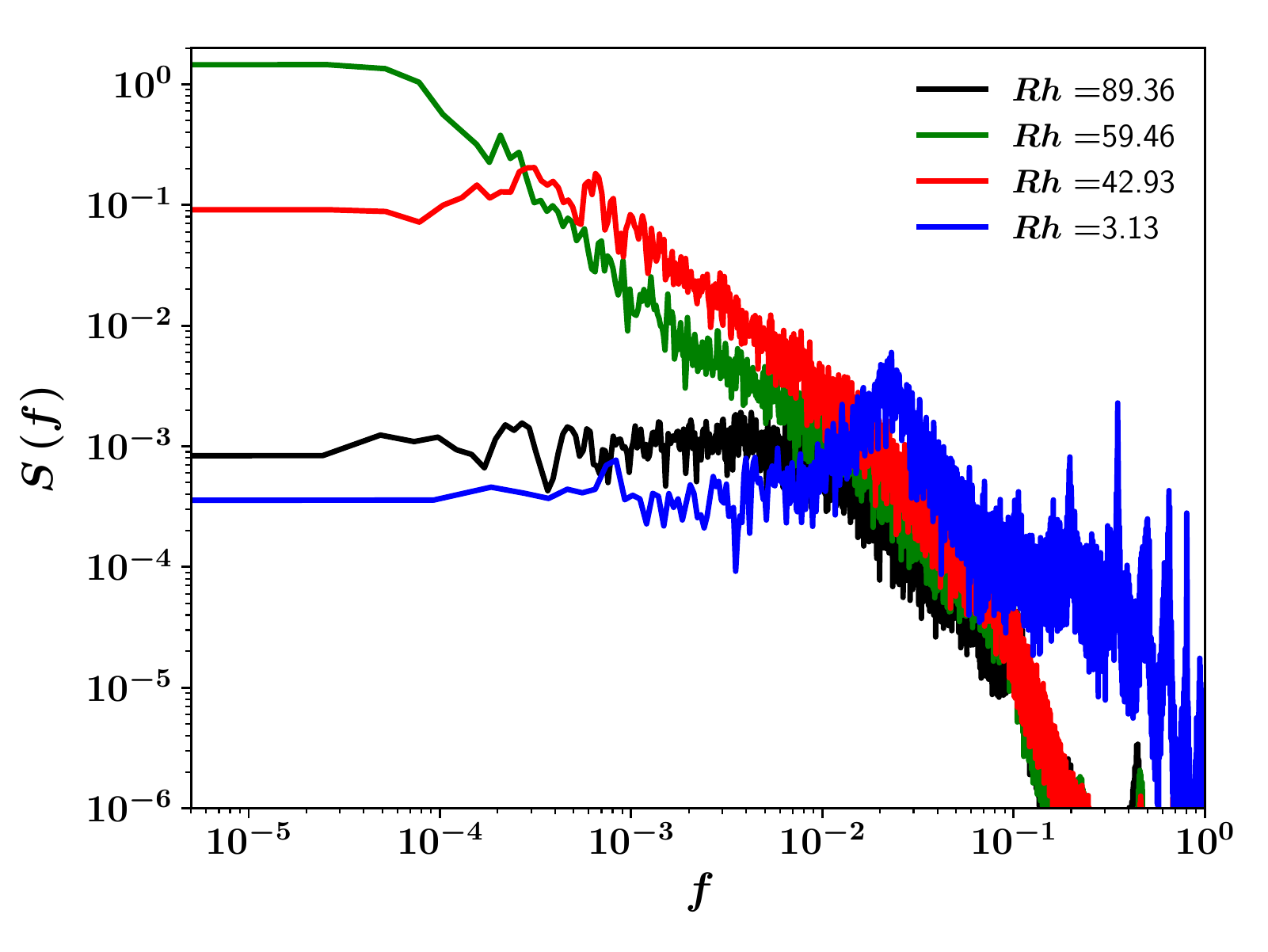}
   \caption{}
   \label{fig:dns_spec_rh}
 \end{subfigure}
 \begin{subfigure}{0.49\textwidth}
   \includegraphics[width=\textwidth]{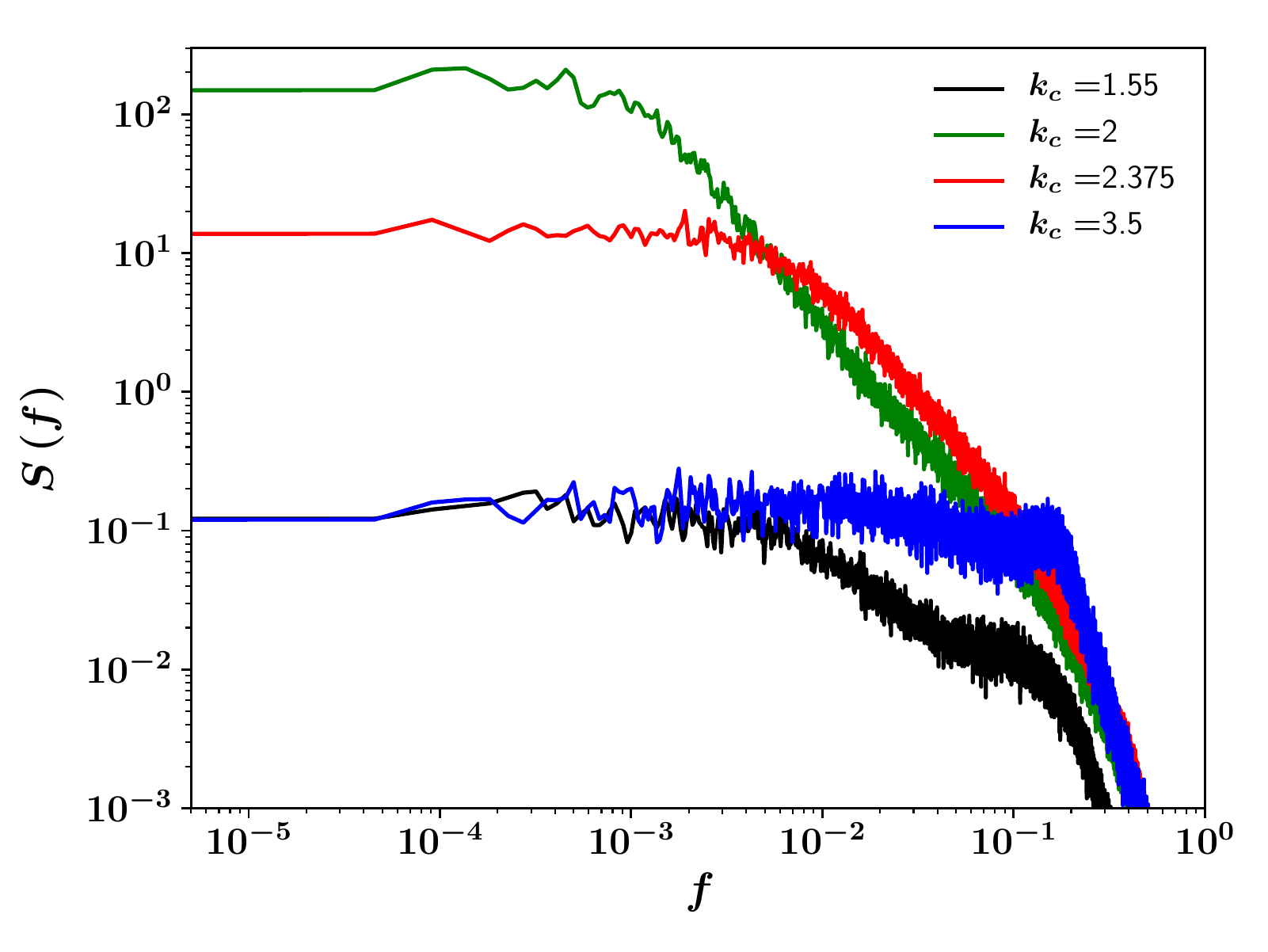}
   \caption{}
   \label{fig:euler_spec_kc}
 \end{subfigure}
 \caption{(Color online) Power spectra of the large scale mode $\lhat\psi_{0,1}$ for different values (a) of $Rh$ for the Navier-Stokes equations and (b) of $k_c$ for the TEE.} %\vd{Kanna: do the PSD of the $Rh = 59.46$ signal for different lengths of the signal}}  
 \label{fig:spec_rh_kc} 
 \end{figure}

To further support our argument about long-time memory we analyse the spectral properties of the large scale mode $\lhat\psi_{0,1}$. 
In Fig. \ref{fig:dns_spec_rh} we plot the power spectra $S(f)$ for different values of $Rh$. These spectra were computed from the time series of the Navier-Stokes equations presented in Fig. \ref{fig:dns_time_series}. 
In the Gaussian regime with $Rh = 3.13$, the spectrum is flat across a large range of frequencies and after a cross-over frequency $f_c$ (where $S(f)$ peaks) it decays. 
In the bimodal regime with $Rh = 42.93$ and $Rh = 59.46$ the largest part of the frequencies display a power law spectrum $S(f) \propto 1/f^\alpha$. As $Rh$ increases within the bimodal regime, the cross-over frequency $f_c$ between $f^0$ and $1/f^\alpha$ moves toward lower frequencies. In other words, the $1/f^\alpha$ range extends to lower frequencies while the $f^0$ range decreases. 
It is interesting to observe that the level of the power spectrum %$1/f^\alpha$-type noise 
decays with the observation time similar to studies of other systems that display $1/f^\alpha$-type noise \cite{sadeghetal14}.
Finally, in the unimodal regime with $Rh = 89.36$ the cross-over frequency $f_c$ between $f^0$ and $1/f^\alpha$ moves to higher frequencies and the amplitude of the $f^0$ part of the spectrum becomes of the same order with the spectrum for $Rh = 3.13$ from the Gaussian regime.

Figure \ref{fig:euler_spec_kc} on the other hand shows the power spectra from the time series of Fig. \ref{fig:euler_time_series}, which correspond to different values of $k_c$ for the TEE minimal model with 15 modes. 
The behaviour we just described for the spectra as $Rh$ increased is reproduced by the TEE minimal model as $k_c$ varies from the Gaussian to the bimodal and then to the unimodal regime (see Fig. \ref{fig:euler_spec_kc}).

Understanding of such low frequency noise is of great importance because power spectra of the type
\begin{equation}
 S(f) \propto 1/f^\alpha 
\end{equation}
with $0 < \alpha < 2$ have been observed in many systems ranging from 
voltage and current fluctuations in vacuum tubes \cite{hoogeetal81},
atmosphere and oceans \cite{fraedrichiblender03,costaetal14}, 
astrophysical magnetic fields \cite{matthaeusgoldstein86} and 
turbulent flows \cite{dmitrukmatthaeus07,dmitruketal14,heraultetal15b,pereiraetal19}. 
These systems often display either fluctuations between symmetric states occurring after random waiting times $\tau$ or an intermittent regime with random asymmetric bursts.
For this kind of processes, it has been shown that the $1/f^\alpha$ spectrum is related to a power-law distribution of the waiting time $\tau$ of the form
\begin{equation}
 \mathcal P(\tau) \propto 1/\tau^{\beta} 
\end{equation}
with the exponents $\alpha$ and $\beta$ satisfying the relation 
$\alpha + \beta = 3$ \cite{lowenteich93}.
This relation is valid for random transitions between symmetric states when {$0 < \alpha < 2$ and $1 < \beta < 3$, \cor{and it also holds for random asymmetric bursts when $0 < \alpha < 1$ and $2 < \beta < 3$ \cite{heraultetal15b}. We see that the large scale mode $\widehat{\psi}_{0,1}$ transitions randomly between two symmetric states with the time average of $\widehat{\psi}_{0,1}$ being zero in the Gaussian and bimodal regimes. We will thus focus on the case $\alpha + \beta = 3$ with $0 < \alpha < 2$ and $1 < \beta < 3$.}

Now we present a simple argument to show how this relation between the exponents of the distribution and the spectrum emerges. \cor{We refer to the following studies where detailed derivations are given \cite{lowenteich93,heraultetal15b,niemann2013fluctuations}. We mention that even though we consider waiting time distributions $P (\tau) \sim \tau^{-\beta}$ with $\beta > 1$, the mean waiting time $\bar{\tau}$ is defined as long as there is an exponential cut-off time scale, see \cite{lowenteich93,niemann2013fluctuations}}. A typical signal of such a process is shown in Fig. \ref{fig:euler_time_series} for $k_c = 2$. From the signal of $\lhat\psi_{0,1}(t)$ we can obtain the auto-correlation 
\begin{equation}
\cor{C(t) = \frac{1}{T \sigma(\widehat{\psi}_{0,1})^2} \int_0^{T} \lhat\psi_{0,1}(s)\lhat\psi_{0,1}(s-t) ds}
\end{equation}
where $T$ is the total duration of the signal \cor{and $\sigma(\widehat{\psi}_{0,1})$ is the standard deviation of the signal}. Looking at the signal of $\lhat\psi_{0,1}(t)$ (see Fig. \ref{fig:euler_time_series}) we observe abrupt fluctuations between the states $\lhat\psi_{0,1}(t) > 0$ and $\lhat\psi_{0,1}(t) < 0$, and phases of long durations $\tau$ of $\lhat\psi_{0,1}(t)$ with the same sign. \cor{Here we assume that the average contributions of short phases do not contribute to the autocorrelation which allows us to use moving average in order to have cleaner signals. The quantity $\widehat{\psi}_{0,1}(s) \widehat{\psi}_{0,1}(s-t)$ integrated during the long 
phases contribute a value proportional to $\tau - t$ over a length $\tau$ for $\tau > t$ and almost zero for $\tau < t$. The time interval $s \in (0, T)$ is then replaced by the variable $\tau \in (t, T)$ and the autocorrelation function \vd{can be approximated as,}}
\begin{equation}
C(t) \approx \frac{1}{T} \int_t^{T} (\tau - t) n(\tau) d\tau \quad \text{for} \quad \overline{\tau} \ll t \ll T
\end{equation}
where $n(\tau)$ is the number of phases of duration $\tau$. This approximation of the autocorrelation $C(t)$ can be written in terms of the PDF of the waiting times $P(\tau)$ between transitions as
\begin{equation}
C(t) \approx \frac{1}{\overline{\tau}} \int_t^{T} (\tau - t) \mathcal P(\tau) d\tau
\end{equation}
because $n(\tau) = \mathcal P(\tau)T/\overline{\tau}$ with $T/\overline{\tau}$ the total number of events.
For $\mathcal P(\tau) \propto 1/\tau^\beta$ we have 
\begin{equation}
 C(t) \propto t^{-\beta+2}
\end{equation}
with $\beta > 2$. Then, using the Wiener-Khinchin theorem we can get the power spectrum $S(f)$ of $\lhat\psi_{0,1}$ by taking the Fourier transform of $C(t)$ as $T \to \infty$, which is
\begin{equation}
 S(f) \propto \frac{1}{f^{3-\beta}}\int_{0}^\infty \omega^{-\beta+2} \cos(\omega) d \omega
\end{equation}
with $\omega = f t$. 
Thus, the relation between the exponent of the spectrum and the exponent of the PDF of waiting times of transitions between the symmetric states $\lhat\psi_{0,1}(t) > 0$ and $\lhat\psi_{0,1}(t) < 0$ is $\alpha = 3-\beta$.

\cor{Figure \ref{fig:dns_spec} shows the power spectrum $S(f)$ of the large scale mode $\lhat\psi_{0,1}$ in the DNS of the Navier-Stokes equations for the case $Rh = 59.45$ and Fig. \ref{fig:euler_spec} the power spectrum for the TEE system for $k_c = 2$.}
 \begin{figure}[!ht]
 \begin{subfigure}{0.49\textwidth}
   \includegraphics[width=\textwidth]{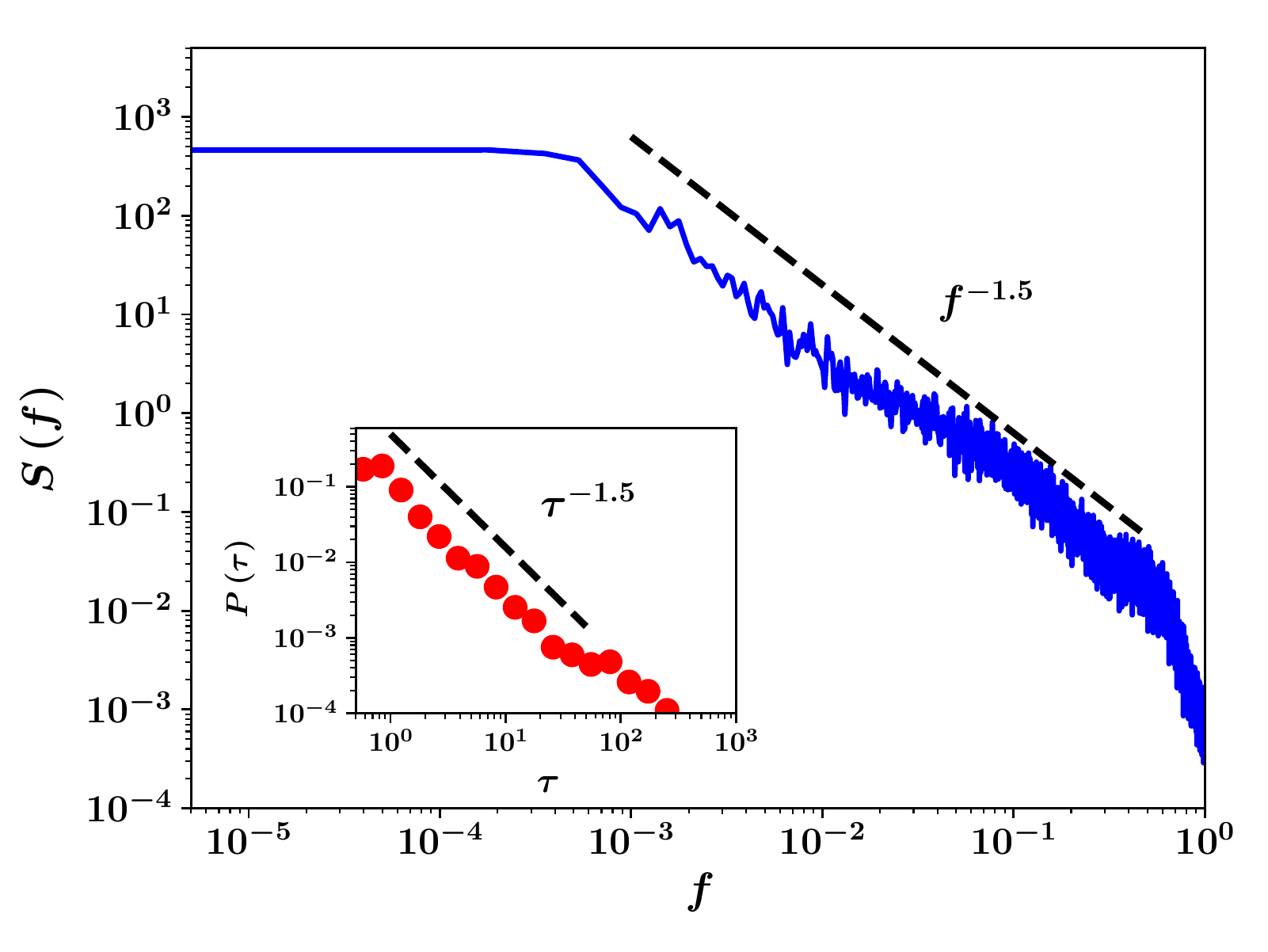}
   \caption{}
   \label{fig:dns_spec}
 \end{subfigure}
 \begin{subfigure}{0.49\textwidth}
   \includegraphics[width=\textwidth]{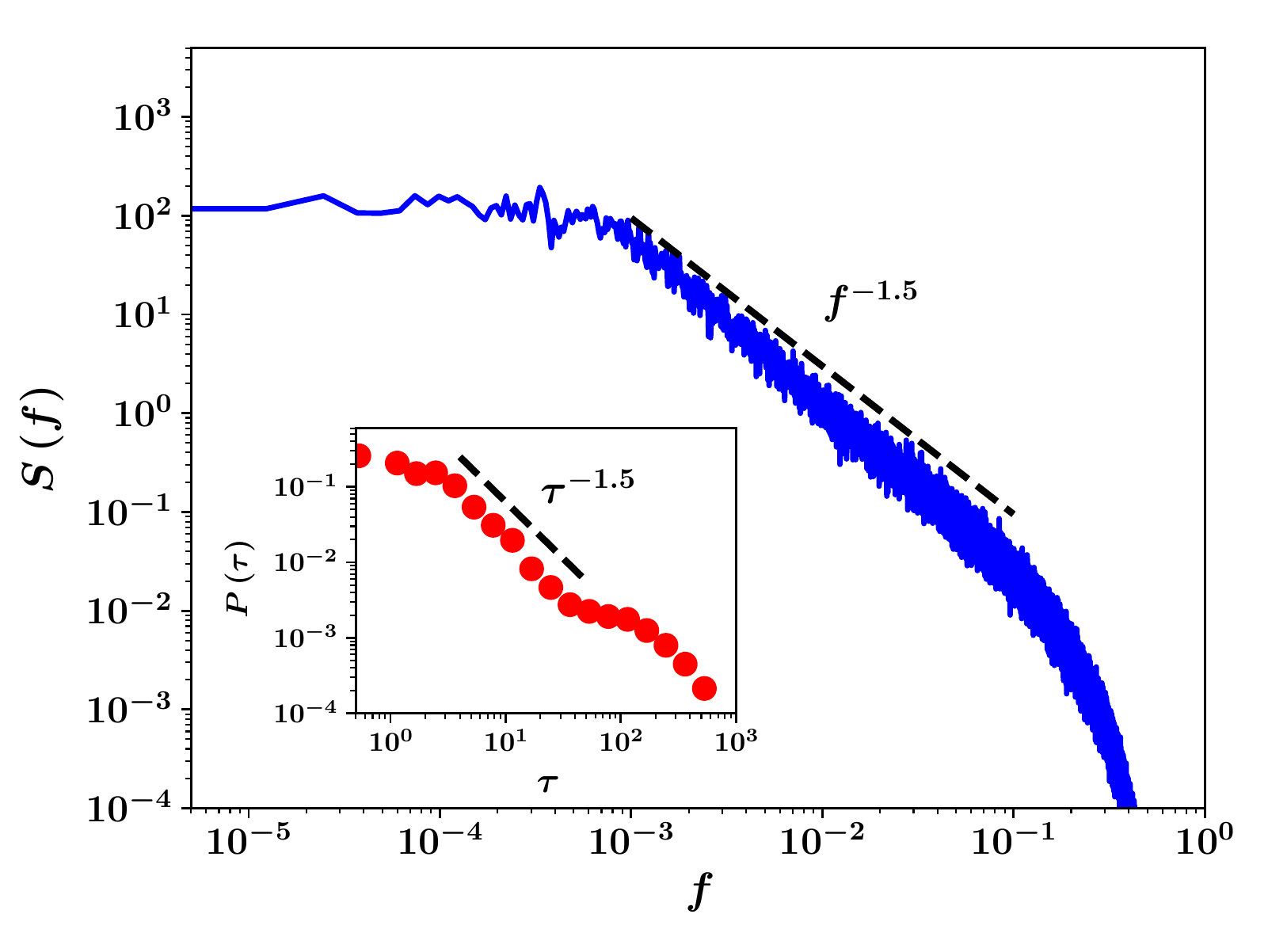}
   \caption{}
   \label{fig:euler_spec}
 \end{subfigure}
 \caption{(Color online) \cor{Power spectrum of the large scale mode $\lhat\psi_{0,1}$ from (a) Navier-Stokes equations for $Rh = 59.45$ and (b) TEE for $k_c = 2$. The insets show the probability density function of the mean waiting time \vd{$P(\tau)$} of successive reversals. The dashed lines show power laws of $f^{-1.5}$ and $\tau^{-1.5}$ (insets) for comparison. } } 
 \label{fig:spec} 
 \end{figure}
\cor{The power spectra from the DNS shows a range of frequencies where $S(f)$ has a scaling of $f^{-1.5}$. We then compare this scaling with the TEE model and we find \vd{here that the scaling is also close to $f^{-1.5}$.}} The $1/f^{1.5}$ noise is between the well understood white noise with $f^0$ power spectrum and the random walk (Brownian motion) noise with $1/f^2$ power spectrum. However, direct differentiation or integration of such convenient signals does not give the required $1/f^{1.5}$ spectrum. 

The process for generating $1/f^\alpha$ noise here differs strongly from low-dimensional dissipative dynamical systems because it involves a large number of degrees of freedom with a large network of triads interacting nonlinearly. In addition, $1/f^\alpha$ noise subsist in the TEE model which represents a system at thermal equilibrium. Our results indicate that the large scale coherent structures of our turbulent flows are responsible for the long-term dynamical memory of the large scale jets and hence the emergence of the $1/f^{1.5}$ noise.

The corresponding PDF of the waiting times $\mathcal P(\tau)$ are shown as insets in Fig. \ref{fig:dns_spec} for the Navier-Stokes equations and in Fig. \ref{fig:euler_spec} for the TEE model. \cor{Both PDFs show a range of time scales in which a power-law distribution of $\mathcal P(\tau) \propto \tau^{-1.5}$ compares well with the data.} These waiting times, distributed as a power law, reflect the scale invariant nature of the statistics, and they are associated with the durations spent by the system in the two states. Such heavy-tailed PDFs indicate that long durations have a high probability of occurrence. The scaling of the spectra and of the distribution of the waiting times are in agreement with the relation $\alpha + \beta = 3$ both for the Navier-Stokes equations and the mininal TEE model. 
%\cor{We finally we mention that the agreement between the exponents of both the TEE and the DNS have been verified only at the qualitative level.}

%
\section{Conclusion}

We have studied in detail the bifurcations of the large scale flow for a turbulent shear flow driven by a Kolmogorov forcing. The domain studied here is anisotropic with an aspect ratio of $2$, and so are the boundary conditions that were taken to be \cor{free-slip in the
\vd{spanwise} direction, and periodic in the \vd{streamwise} direction}. The geometry and the forcing leads to the formation of jets, with the largest mode in the system $\widehat{\psi}_{0,1}$ corresponding to two counter propagating jets. The mode $\widehat{\psi}_{0,1}$ is not directly excited by the forcing but the energy injected into the forcing scale is transferred to the largest mode $\widehat{\psi}_{0,1}$ by an inverse nonlinear transfer. 
A non-zero value of $\widehat{\psi}_{0,1}$ does not respect the centreline symmetry $y = \pi L_y /2$. 

With the friction Reynonds number $Rh$ as the control parameter in the system, we see that the mode $\widehat{\psi}_{0,1}$ first displays a Gaussian behaviour in the turbulent regime. Then, as we increase $Rh$, the large scale mode transitions to a bimodal behaviour denoting the first bifurcation. For larger $Rh$, we get to another bifurcation with a unimodal distribution where the $\widehat{\psi}_{0,1}$ mode no longer reverses for the whole duration of the simulation. 
\cor{These bifurcations of the large scale mode $\widehat{\psi}_{0,1}$ are related to changes in the form of the time averaged mean flow $\left\langle U(y, t) \right\rangle_t$. Similar phenomena in the mean flow profile are also observed in the drag crisis that is a well known example of transition of a fully turbulent flow \cite{tritton77}.}
Using the truncated Euler equation we are able to construct a minimal model, which consists of $15$ modes, that effectively captures these bifurcations which occur on top of the turbulent background flow. 
The control parameter in the TEE model is the ratio of the enstrophy and energy at time $t=0$ defined as $k_c^2 = \Omega/E$. The role of $Rh$ as the control parameter for the Navier-Stokes equations is now replaced by $k_c$ for the TEE model.

We also compared the power spectra from the time series of $\widehat{\psi}_{0,1}$ and the PDF of the mean waiting times $\tau$ between the two symmetric turbulent states for both the Navier-Stokes and the TEE systems. \vd{From the Navier-Stokes equations} we found the power spectra to scale as $S(f) \propto 1/f^{1.5}$ and the PDFs of the mean waiting times to obey a power law $\mathcal P(\tau) \propto 1/\tau^{1.5}$ in the bimodal regime. 
%\cor{The numerical solutions of the TEE model reproduces qualitatively the scaling laws of the power spectrum and PDF of the mean waiting time. 
\cor{Moreover, we showed numerically that a minimal TEE model, which is a system at thermal equilibrium, can qualitatively reproduce these scaling laws for the $1/f^{\alpha}$ noise and the PDF of the mean waiting time. This is an interesting result since it has been sometimes believed that $1/f^{\alpha}$ noise only occurs in systems out of thermal equilibrium \cite{duttahorn81}. Although analytical results on TEE exist both on the canonical and microcanonical ensembles, they are limited so far to spatial spectra. It would be of interest to find similar results on temporal spectra or equivalently on correlation functions at different times and to compare them with DNS.}
 
% In the broader context of understanding the behaviour of zonal jet like structures in atmospheric and oceanic flows, the minimal model from the TEE seems to have all the necessary ingredients to capture the bifurcations of the large scale mode
 
\cor{In this paper we have provided a new example where the truncated Euler equations capture the bifurcations of the large scale flow between turbulent states. In this case, the TEE model can capture the bifurcations of  the large scale mode even if it is not orders of magnitude above the rest of the modes like in previous studies, a regime where mean field approach or asymptotics might not be applicable.
Future studies \vd{should} aim at understanding whether the TEE model} \vd{can reproduce transitions of the large scale dynamics in turbulent flows from more realistic models for the dynamics of planetary atmospheres and the ocean, such as advanced quasi-geostrophic and shallow water models.}

\section*{Acknowledgements}
The authors would like to thank F. P\'etr\'elis and V. Shukla for useful discussions in the early stage of this work.

%\pagebreak 
 
\bibliographystyle{unsrt}
\bibliography{references}

\begin{thebibliography}{10}

\bibitem{dijkstra13}
H.~A. Dijkstra.
\newblock {\em Nonlinear climate dynamics}.
\newblock Cambridge University Press, 2013.

\bibitem{ghil00}
M.~Ghil.
\newblock Is our climate stable? bifurcations, transitions and oscillations in
  climate dynamics.
\newblock {\em Science for survival and sustainable development}, pages
  163--184, 2000.

\bibitem{stockeretal13}
T.~F. Stocker, D.~Qin, G.-K. Plattner, M.~Tignor, S.~K. Allen, J.~Boschung,
  A.~Nauels, Y.~Xia, V.~Bex, P.~M. Midgley, et~al.
\newblock Climate change 2013: The physical science basis, 2013.

\bibitem{itohkimoto96}
H.~Itoh and M.~Kimoto.
\newblock Multiple attractors and chaotic itinerancy in a quasigeostrophic
  model with realistic topography: Implications for weather regimes and
  low-frequency variability.
\newblock {\em J. Atmos. Sci.}, 53(15):2217--2231, 1996.

\bibitem{schmeitsdijkstra01}
M.~J. Schmeits and H.~A. Dijkstra.
\newblock Bimodal behavior of the kuroshio and the gulf stream.
\newblock {\em J. Phys. Oceanogr.}, 31(12):3435--3456, 2001.

\bibitem{weeksetal97}
E.~R. Weeks, Y.~Tian, J.~S. Urbach, K.~Ide, H.~L. Swinney, and M.~Ghil.
\newblock Transitions between blocked and zonal flows in a rotating annulus
  with topography.
\newblock {\em Science}, 278(5343):1598--1601, 1997.

\bibitem{seminetal18}
B.~Semin, N.~Garroum, F.~P{\'e}tr{\'e}lis, and S.~Fauve.
\newblock Nonlinear saturation of the large scale flow in a laboratory model of
  the quasibiennial oscillation.
\newblock {\em Phys. Rev. Lett.}, 121:134502, 2018.

\bibitem{renaud2019periodicity}
A.~Renaud, L.-P. Nadeau, and A.~Venaille.
\newblock Periodicity disruption of a model quasibiennial oscillation of
  equatorial winds.
\newblock {\em Phys. Rev. Lett.}, 122(21):214504, 2019.

\bibitem{wygnanskietal86}
I~Wygnanski, F~Champagne, and B~Marasli.
\newblock On the large-scale structures in two-dimensional, small-deficit,
  turbulent wakes.
\newblock {\em J. Fluid Mech.}, 168:31--71, 1986.

\bibitem{cadotetal15}
O.~Cadot, A.~Evrard, and L.~Pastur.
\newblock Imperfect supercritical bifurcation in a three-dimensional turbulent
  wake.
\newblock {\em Phys. Rev. E}, 91:063005, 2015.

\bibitem{breuerhansen09}
M.~Breuer and U.~Hansen.
\newblock Turbulent convection in the zero reynolds number limit.
\newblock {\em {EPL} (Europhysics Letters)}, 86:24004, 2009.

\bibitem{stevensetal09}
R.~J. A.~M. Stevens, J.-Q. Zhong, H.~J.~H. Clercx, G.~Ahlers, and D.~Lohse.
\newblock Transitions between turbulent states in rotating
  {R}ayleigh-{B\'e}nard convection.
\newblock {\em Phys. Rev. Lett.}, 103:024503, 2009.

\bibitem{sugiyamaetal10}
Kazuyasu Sugiyama, Rui Ni, Richard J. A.~M. Stevens, Tak~Shing Chan, Sheng-Qi
  Zhou, Heng-Dong Xi, Chao Sun, Siegfried Grossmann, Ke-Qing Xia, and Detlef
  Lohse.
\newblock Flow reversals in thermally driven turbulence.
\newblock {\em Phys. Rev. Lett.}, 105:034503, 2010.

\bibitem{chandramahendra13}
M.~Chandra and M.~K. Verma.
\newblock Flow reversals in turbulent convection via vortex reconnections.
\newblock {\em Phys. Rev. Lett.}, 110:114503, 2013.

\bibitem{labbeetal96}
R.~Labb{\'e}, J.‐F. Pinton, and S.~Fauve.
\newblock Study of the von k{\'a}rm{\'a}n flow between coaxial corotating
  disks.
\newblock {\em Phys. Fluids}, 8(4):914--922, 1996.

\bibitem{raveletetal04}
F.~Ravelet, L.~Mari{\'e}, A.~Chiffaudel, and F.~Daviaud.
\newblock Multistability and memory effect in a highly turbulent flow:
  Experimental evidence for a global bifurcation.
\newblock {\em Phys. Rev. Lett.}, 93(16):164501, 2004.

\bibitem{torreburguete07}
A.~de~la Torre and J.~Burguete.
\newblock Slow dynamics in a turbulent von {K\'a}rm{\'a}n swirling flow.
\newblock {\em Phys. Rev. Lett.}, 99:054101, 2007.

\bibitem{berhanuetal07}
M.~Berhanu, R.~Monchaux, S.~Fauve, N.~Mordant, F.~P{\'e}tr{\'e}lis,
  A.~Chiffaudel, F.~Daviaud, B.~Dubrulle, L.~Mari{\'e}, F.~Ravelet, et~al.
\newblock Magnetic field reversals in an experimental turbulent dynamo.
\newblock {\em Europhys. Lett.}, 77(5):59001, 2007.

\bibitem{micheletal16}
G.~Michel, J.~Herault, F.~P{\'{e}}tr{\'{e}}lis, and S.~Fauve.
\newblock Bifurcations of a large-scale circulation in a quasi-bidimensional
  turbulent flow.
\newblock {\em Europhys. Lett.}, 115(6):64004, 2016.

\bibitem{mishraetal15}
P.~K. Mishra, J.~Herault, S.~Fauve, and M.~K. Verma.
\newblock Dynamics of reversals and condensates in two-dimensional kolmogorov
  flows.
\newblock {\em Phys. Rev. E}, 91(5):053005, 2015.

\bibitem{zeitlin18}
Vladimir Zeitlin.
\newblock {\em Geophysical fluid dynamics: understanding (almost) everything
  with rotating shallow water models}.
\newblock Oxford University Press, 2018.

\bibitem{bouchetsimonnet09}
F.~Bouchet and E.~Simonnet.
\newblock Random changes of flow topology in two-dimensional and geophysical
  turbulence.
\newblock {\em Phys. Rev. Lett.}, 102(9):094504, 2009.

\bibitem{galperinread19}
B.~Galperin and P.~L. Read.
\newblock {\em Zonal Jets: Phenomenology, Genesis, and Physics}.
\newblock Cambridge University Press, 2019.

\bibitem{bouchetetal19}
F.~Bouchet, J.~Rolland, and E.~Simonnet.
\newblock Rare event algorithm links transitions in turbulent flows with
  activated nucleations.
\newblock {\em Phys. Rev. Lett.}, 122(7):074502, 2019.

\bibitem{kadanoff00}
L.~P. Kadanoff.
\newblock {\em Statistical physics: statics, dynamics and renormalization}.
\newblock World Scientific Publishing Company, 2000.

\bibitem{shuklaetal16}
V.~Shukla, S.~Fauve, and M.-E. Brachet.
\newblock Statistical theory of reversals in two-dimensional confined turbulent
  flows.
\newblock {\em Phys. Rev. E}, 94(6):061101, 2016.

\bibitem{kraichnan67}
R.~H. Kraichnan.
\newblock Inertial ranges in two‐dimensional turbulence.
\newblock {\em Phys. Fluids}, 10(7):1417--1423, 1967.

\bibitem{kraichnan75}
R.~H. Kraichnan.
\newblock Statistical dynamics of two-dimensional flow.
\newblock {\em J. Fluid Mech.}, 67(1):155–175, 1975.

\bibitem{gottlieborszag77}
David Gottlieb and Steven~A Orszag.
\newblock {\em Numerical analysis of spectral methods: theory and
  applications}, volume~26.
\newblock Siam, 1977.

\bibitem{thess92}
A.~Thess.
\newblock Instabilities in two-dimensional spatially periodic flows. part i:
  Kolmogorov flow.
\newblock {\em Phys. Fluids A}, 4(7):1385--1395, 1992.

\bibitem{sdf19b}
K.~Seshasayanan, V.~Dallas, and S.~Fauve.
\newblock Bifurcations of a plane parallel flow with kolmogorov forcing.
\newblock {\em arXiv preprint arXiv:2004.12418}, 2020.

\bibitem{fauveetal17}
S.~Fauve, J.~Herault, G.~Michel, and F.~P{\'e}tr{\'e}lis.
\newblock Instabilities on a turbulent background.
\newblock {\em J. Stat. Mech.}, 2017(6):064001, 2017.

\bibitem{lee52}
T.~D. Lee.
\newblock On some statistical properties of hydrodynamical and
  magneto-hydrodynamical fields.
\newblock {\em Quart. Appl. Math.}, 10(1):69--74, 1952.

\bibitem{khinchin60}
A.~I. Khinchin.
\newblock {\em Mathematical foundations of statistical mechanics}.
\newblock Dover Publications, 1960.

\bibitem{dfa15}
V.~Dallas, S.~Fauve, and A.~Alexakis.
\newblock Statistical equilibria of large scales in dissipative hydrodynamic
  turbulence.
\newblock {\em Phys. Rev. Lett.}, 115:204501, 2015.

\bibitem{ld16}
M.~Linkmann and V.~Dallas.
\newblock Large-scale dynamics of magnetic helicity.
\newblock {\em Phys. Rev. E}, 94:053209, 2016.

\bibitem{alexakisbrachet19}
A.~Alexakis and M.-E. Brachet.
\newblock On the thermal equilibrium state of large-scale flows.
\newblock {\em J. Fluid Mech.}, 872:594–625, 2019.

\bibitem{heraultetal15b}
J.~Herault, F.~P{\'e}tr{\'e}lis, and S.~Fauve.
\newblock $1/f^{\alpha}$ {L}ow frequency fluctuations in turbulent flows.
\newblock {\em J. Stat. Phys.}, 161(6):1379--1389, 2015.

\bibitem{pereiraetal19}
M.~Pereira, C.~Gissinger, and S.~Fauve.
\newblock 1/f noise and long-term memory of coherent structures in a turbulent
  shear flow.
\newblock {\em Phys. Rev. E}, 99(2):023106, 2019.

\bibitem{sadeghetal14}
A.~Sadegh, E.~Barkai, and D.~Krapf.
\newblock 1/f noise for intermittent quantum dots exhibits non-stationarity and
  critical exponents.
\newblock {\em New J. Phys.}, 16(11):113054, 2014.

\bibitem{hoogeetal81}
F.~N. Hooge, T.~G.~M. Kleinpenning, and L.~K.~J. Vandamme.
\newblock Experimental studies on 1/f noise.
\newblock {\em Rep. Prog. Phys.}, 44(5):479--532, 1981.

\bibitem{fraedrichiblender03}
K.~Fraedrich and R.~Blender.
\newblock Scaling of atmosphere and ocean temperature correlations in
  observations and climate models.
\newblock {\em Phys. Rev. Lett.}, 90:108501, 2003.

\bibitem{costaetal14}
A.~Costa, A.~R. Osborne, D.~T. Resio, S.~Alessio, E.~Chriv\`{\i}, E.~Saggese,
  K.~Bellomo, and C.~E. Long.
\newblock Soliton turbulence in shallow water ocean surface waves.
\newblock {\em Phys. Rev. Lett.}, 113:108501, 2014.

\bibitem{matthaeusgoldstein86}
W.~H. Matthaeus and M.~L. Goldstein.
\newblock Low-frequency $\frac{1}{f}$ noise in the interplanetary magnetic
  field.
\newblock {\em Phys. Rev. Lett.}, 57:495--498, 1986.

\bibitem{dmitrukmatthaeus07}
P.~Dmitruk and W.~H. Matthaeus.
\newblock Low-frequency 1/f fluctuations in hydrodynamic and
  magnetohydrodynamic turbulence.
\newblock {\em Phys. Rev. E}, 76:036305, 2007.

\bibitem{dmitruketal14}
P.~Dmitruk, P.~D. Mininni, A.~Pouquet, S.~Servidio, and W.~H. Matthaeus.
\newblock Magnetic field reversals and long-time memory in conducting flows.
\newblock {\em Phys. Rev. E}, 90:043010, 2014.

\bibitem{lowenteich93}
S.~B. Lowen and M.~C. Teich.
\newblock Fractal renewal processes generate 1/f noise.
\newblock {\em Phys. Rev. E}, 47:992--1001, 1993.

\bibitem{niemann2013fluctuations}
M.~Niemann, H.~Kantz, and E.~Barkai.
\newblock Fluctuations of 1/f noise and the low-frequency cutoff paradox.
\newblock {\em Phys. Rev. Lett.}, 110(14):140603, 2013.

\bibitem{tritton77}
D.~J. Tritton.
\newblock {\em Physical Fluid Dynamics}.
\newblock Springer Netherlands, 1977.

\bibitem{duttahorn81}
P.~Dutta and P.~M. Horn.
\newblock Low-frequency fluctuations in solids: $\frac{1}{f}$ noise.
\newblock {\em Rev. Mod. Phys.}, 53:497--516, 1981.

\end{thebibliography}

\end{document}